\documentclass[12pt,preprint]{aastex}
\usepackage{amsmath}

\newcommand{\degK}{\,{}^\circ\mbox{K}}

\newcommand{\gr}{\mbox{\,g}}
\newcommand{\cm}{\mbox{\,cm}}
\newcommand{\meter}{\mbox{\,m}}
\newcommand{\au}{\mbox{\,AU}}

\newcommand{\sech}{\mbox{\,sech}}
\newcommand{\eff}{\mathit{eff}}

\newcommand{\vect}[1]{\mbox{\boldmath$#1$}}
\newcommand{\me}{\mathrm{e}}
\newcommand{\mi}{\mathrm{i}}
\newcommand{\dif}{\mathrm{d}}
\newcommand{\Ri}{\mathrm{Ri}}

\begin{document}

\title{The effect of the Coriolis force on Kelvin-Helmholtz-driven
mixing in protoplanetary disks.}

\author{Gilberto C. G\'omez}
\email{gomez@astro.umd.edu}

\and

\author{Eve C. Ostriker}
\email{ostriker@astro.umd.edu}

\affil{Department of Astronomy, University of Maryland,
  College Park, MD 20742, U.S.A.}


\begin{abstract}

We study the stability of proto-planetary disks with vertical
velocity gradients in their equilibrium rotation rates; such
gradients are expected to develop when dust settles into the
midplane.
Using a linear stability analysis of a simple three-layer
model, we show that the onset of instability occurs at
a larger value of the Richardson number, and therefore for a thicker
layer, when the ef\-fects of Coriolis forces are included.
This analysis also shows that even-symmetry (midplane-crossing)
modes develop faster than odd-symmetry ones.
These conclusions are corroborated by a large number of nonlinear
numerical simulations with two different parameterized prescriptions
for the initial (continuous) dust distributions.
Based on these numerical experiments, the Richardson number required 
for marginal stability is more than an order of magnitude
larger than the traditional $1/4$ value.
The dominant modes that grow have horizontal wavelengths of several
initial dust scale heights,
and in  nonlinear stages mix solids fairly homogeneously over
a comparable vertical range.
We conclude that gravitational instability may be more difficult to
achieve than previously thought, and that the vertical distribution
of matter within the dust layer is likely globally, rather than
locally, determined.

\end{abstract}

\keywords{hydrodynamics ---
instabilities ---
turbulence ---
planets and satellites: formation ---
planetary systems: protoplanetary disks}

\section{INTRODUCTION}
\label{intro_sec}

In the generally accepted model for the formation of planets, the
circumstellar disk -- consisting of material remaining from the
formation of the central star -- progresses through a series of
dynamical stages.
Over the course of this evolution, the solids
contained in the disk are believed to settle towards the midplane and
form planetesimals, which in turn merge to assemble terrestrial
planets and cores of gaseous giants.
Unfortunately, some of the most basic elements of this model are
not understood, with the detailed physical processes involved at many
stages only barely explored.

One such poorly-understood stage is the formation of planetesimals from
interstellar dust.%
\footnote{
  In the outer parts of the disk, gaseous water, ammonia, and other
  molecules condense into ice, augmenting the dust with additional
  solids.
}
It is generally accepted that micron-sized particles collide and
stick together to form larger particles,
but it is not clear that this stochastic coagulation
process can continue up the ladder to generate km-sized planetesimals.
Among other difficulties with this model is the problem that
bodies with approximate sizes between $1\cm$ and $1\meter$ would
drift radially inward very fast, as a
consequence of drag with the partially pressure-supported gas disk
in which they are embeded \citep{wei77}.
As a way of solving this problem, authors have looked for
a process that allows particles to grow through this range of sizes
in time scales similar or smaller to the orbital time
\citep{gol73, wei80, nak81, nak86, wei93}.
For example, \citet{cuz93} and \citet{cha95} noted that because
particles of different sizes are affected differently by gas drag
they would drift at different speeds, thus increasing their collision
rate and possibly increasing the particle growth rate.
Nevertheless, it remains questionable whether solid state forces
can grow particles through sticking at the collision velocities
likely present in proto-planetary disks
(see discussion in Youdin \& Shu 2002).

Gravitational instability (GI) has been proposed as an alternative
\citep{saf69, gol73, cor81, sek83}.
In this model, as the dust settles vertically to create
a dense midplane subdisk, the spatial particle density may increase
until the layer crosses the GI density threshold.
This model has the advantage that particle growth from mm
to km sizes happens in a short time scale.
Still, there are uncertainties in this GI model.
For example,
several authors \citep{wei93, cuz94, wei95, tan04} have pointed out
that $\sim 10 \meter$ sized bodies, an intermediate step of the
two-staged process proposed by \citet{gol73}, reach the disk midplane
with random velocities too large for GI to proceed.

A more frequently discussed problem with the GI hypothesis
arises from the fact that,
as the solid particle density increases through vertical settling,
the gas/dust mixture becomes less sensitive to large scale gas
pressure gradients that are likely to be present in the disk.
With the loss of radially-supporting pressure gradients,
the dust-rich mixture near the midplane would orbit the central
star with an orbital
frequency closer to the keplerian rate, and thus develop a velocity
shear with its dust-poor envelope.
This velocity shear might
trigger the Kelvin-Helmholtz instability (KHI),
which could mix the dust and gas layers and stop particle
settling before GI sets in \citep{gol73, wei80}.
The question posed at this point is: what are the appropriate
circumstances for the development of the KHI in circumstellar disks?
Or conversely, is it possible to have a KH-stable disk
that is susceptible to GI, and so able to form planetesimals
within a very short timescale?

The most favorable circumstance for GI is if there is no turbulence
-- and no mixing of solids -- in the absense of KHI.
\citet{sek00, sek01} performed linear stability analyses
of vertically shearing disks and concluded that GI can be achieved,
avoiding the KHI induced mixing, only if the dust-to-gas
surface density ratio is much larger than the solar abundance.
Other authors
have reached a similar conclusion while considering the potentially
de-stabilizing influence of
the Coriolis and tidal forces \citep{ish02, ish03},
stratification of the medium, two-fluid effects and radiative
cooling \citep{gar04}.
\citet{you02} further argued that the gas involved in the KHI should
be able to drag and mix only a finite amount of solids, so that
there might be a decoupling and precipitation of solids once the
dust space density is large enough,
even for particle sizes at which a good velocity coupling is expected.
The precipitated particles would
be free to continue settling and undergo GI.
If a strong central enhancement of dust develops \citep{sek98}
and precipitates out an
independent solid layer, then potentially this reduces the needed
dust-to-gas surface density enhancement required for planetesimals
to form via GI.
Even with the modification proposed by \citet{you02}, however, the
dust-to-gas surface density ratio must be enhanced above
solar-abundance values $\sim 0.01$ by an order of magnitude.

Of course, this increased surface
density ratio (which can also help accelerate stochastic
coagulation) does not have to be global.
Particle drift in the radial direction
\citep{cuz93, you02, hag03, wei03, you04},
and self-gravitating spiral modes \citep{ric04} could provide
localized surface density concentrations of dust.
Provided that these enhancements are long-lived, this could
lead to localized GI.

As discussed above, when dust settles toward the midplane,
both (de-stabilizing) shear
and (stabilizing) buoy\-an\-cy increase.
In the situation where dust and gas are well-enough coupled, they
may be considered a single fluid.
For a stratified flow, competition between the
opposing shear and buoyancy effects is traditionally
described by the Richardson number,

\begin{equation}
\Ri = \frac{-g ~ \dif \ln \rho/\dif z}{(\dif v/\dif z)^2},
\label{ri_eq}
\end{equation}

\noindent
where $v(z)$ is the equilibrium shear velocity in the plane
perpendicular to $\vect{\hat e}_z$.
\citet{mil61} and \citet{how61} proved (for incompressible flow
in the Boussinesque approximation) that $\Ri<1/4$ somewhere in
the layer is a necessary (but not sufficient) condition for
instability (see also Drazin \& Reid 1981, p. 325;
a review of the proof is presented in the Appendix B of
Li, Goodman \& Narayan 2003).
As \citet{gar04} argue, however, some of the assumptions in
this proof, while acceptable in other physical regimes, are not
applicable to the problem at hand.

In this paper, we explore the influence of an effect not considered
in arriving at the ``$\Ri > 1/4$'' stability criterion,
namely the Coriolis force.
In \S \ref{linear_sec} we present the basic equations of the model
problem we have defined.
In \S \ref{threelayer_sec} we perform a linear stability analysis of
a discrete three-layer configuration, consisting of a dust-rich layer
surrounded by dust-free gas.
This analysis relaxes some of the assumptions made by previous work,
compares modes of instability with even and odd symmetry, and
assesses the effect of the Coriolis force.
In \S \ref{numerics_sec} we present the results of our numerical
experiments, focusing on the effect the Coriolis force has on the
stability of model disks with two different (continuous) initial dust
distributions.
Finally, in \S \ref{conclusions_sec} we present our conclusions.


\section{BASIC EQUATIONS}
\label{linear_sec}

Consider the reference frame of a fluid orbiting a central star at a
distance $R_0$, with angular velocity $\vect{\Omega}_F$.
For adiabatic flow of a gamma-law gas,
the equations of hydrodynamics in this frame read

\begin{eqnarray}
\frac{\partial \rho}{\partial t}
  + \nabla \cdot (\rho \vect{v}) &=& 0 \label{hd_eq1} \\
\frac{\partial \vect{v}}{\partial t}
  + (\vect{v} \cdot \nabla) \vect{v} &=&
  - \frac{\nabla P}{\rho} - \nabla \Phi
  - 2 \vect{\Omega}_F \times \vect{v}
  - \vect{\Omega}_F \times (\vect{\Omega}_F \times\vect{R})
  \label{hd_eq2} \\
\frac{\partial {\cal E}}{\partial t}
  + (\vect{v} \cdot \nabla) {\cal E} &=&
  - ( P + {\cal E}) \nabla \cdot \vect{v} \label{hd_eq3} \\
P &=& (\gamma -1) {\cal E}. \label{hd_eq4}
\end{eqnarray}

\noindent
All the symbols are standard.
We orient the coordinate system with
$\vect{\hat e}_z$ parallel to $\vect{\Omega}_F$,
$\vect{\hat e}_x$ pointing radially,
and $\vect{\hat e}_y$ pointing along the azimuthal background
flow.
We shall treat a small domain compared to $R_0$, and
disregard the curvature terms in equations
(\ref{hd_eq1})-(\ref{hd_eq4}).

In addition to gas with density $\rho_g$, the flow contains a
component of solids (dust + ice).
We asssume that the gas and solids are strongly coupled, so
$\vect{v}_d = \vect{v}_g = \vect{v}$.
In the strongly coupled limit,
the dust component of the mixture adds inertia to the flow so that
$\rho = \rho_g + \rho_d$ in equations (\ref{hd_eq1}) and (\ref{hd_eq2})
but the solids do not contribute to the
pressure ($P = c_s^2 \rho_g$, where $c_s$ is the isothermal sound
speed) and energy density in equations (\ref{hd_eq3}) and
(\ref{hd_eq4}).
The dust component also separately obeys a continuity equation

\begin{equation}
\frac{\partial \rho_d}{\partial t}
  + \nabla \cdot ( \rho_d \vect{v} ) = 0.
\label{dust_eq}
\end{equation}

For an axisymmetric equilibrium,
the $x$ and $z$ components of the momentum equation read

\begin{eqnarray}
\frac{v_{0y}^2(z)}{R_0}
  + 2 \Omega_F v_{0y}(z) + \Omega_F^2 R_0
  - \frac{\partial P_0 / \partial R |_{R_0}}{\rho_d(z)+\rho_g(z)}
  - \Omega_K^2 R_0 &=& 0 \label{momr_eq} \\
- \frac{1}{\rho_d(z)+\rho_g(z)} \frac{\partial P_0}{\partial z}
  - \Omega_K^2 z &=& 0. \label{momz_eq}
\end{eqnarray}

\noindent
Here,
$\partial \Phi / \partial R \approx \Omega_K^2 R$,
where $\Omega_K$ is
the keplerian orbital frequency, and we approximate
$\partial \Phi / \partial z \approx \Omega_K^2 z$.

From equation (\ref{momr_eq}),
the orbital velocity of the flow in equilibrium
depends on the dust abundance.
In order to further specify our coordinate system, let us define
$\Omega_F$ as the inertial-frame orbital frequency of dust-free
gas with a reference value of density at $z=0$ of $\rho_{g0}$.
With this definition of $\Omega_F$, $v_{0y}=0$ when $\rho_d=0$
and we can relate $\Omega_F$ to the
physical parameters of the system,

\begin{equation}
\frac{\Omega_F}{\Omega_K}
  = \left[ 1 + \frac{\partial P_0 / \partial R |_{R_0,z=0}}
                    {\Omega_K^2 R_0 \rho_{g0} } \right]^{1/2}.
\label{omegaf_eq}
\end{equation}

\noindent
Here, $P_0$ is the pressure in the dust-free case.

Solving equation (\ref{momr_eq}) for $v_{0y}$ and taking
$|v_{0y}| \ll \Omega_F R_0$, we obtain

\begin{equation}
v_{0y} = v_{0,max} \left( \frac{\mu}{1+\mu} \right),
\label{vy0_eq}
\end{equation}

\noindent
where $\mu(z) = \rho_d(z) / \rho_g(z)$ is the dust abundance, and

\begin{equation}
v_{0,max} = - \frac{\partial P_0 / \partial R |_{R_0,z=0}}
                 {2 \Omega_F \rho_{g0}}
\label{v0max_eq}
\end{equation}

\noindent
is the maximum speed that the dust-laden gas can attain,
i.~e. the difference between the keplerian velocity and
$\Omega_F R_0$.%
\footnote{
$v_{0,max}$ relates to the $\eta_K$ parameter used in previous
work by $v_{0,max}=\eta_K R_0 \Omega_K^2/\Omega_F$.}

In order to define $P_0(R)$ we will adopt the parametrization of
\citet{you02},

\begin{eqnarray}
\Sigma_g(R)
  &=& 1700 \gr \cm^{-2} ~f_g     \left(\frac{R}{1 \au}\right)^{-p} \\
T(R)
  &=& 280\degK ~f_T \left(\frac{R}{1 \au}\right)^{-q}
\end{eqnarray}

\noindent
where $\Sigma_g$ is the gas surface density, and $T$ is the
temperature of the gas.
For the minimum solar nebula (MSN) model of \citet{hay81},
$f_g=f_T=1$, $p=3/2$ and $q=1/2$.
In this case, $v_{0,max}/c_s$ depends only weakly on $R$,

\begin{equation}
\frac{v_{0,max}}{c_s} = 7.3 \times 10^{-2}
  \, \left(\frac{R}{1 \au}\right)^{1/4}.
\label{v0maxR_eq}
\end{equation}

Consider now the vertical hydrostatic equilibrium.
For vertically-isothermal conditions,
equation (\ref{momz_eq}) can be written as

\begin{equation}
\frac{\partial \ln \rho_g}{\partial z}
  = - \frac{\Omega_K^2 z}{c_s^2} [1 + \mu(z) ],
\label{rhog_eq}
\end{equation}

\noindent
which can be solved (analytically or
numerically) for $\rho_g(z)$, given $\mu(z)$.
The total density is then $\rho = \rho_g (1+\mu)$.
In \S \ref{numerics_sec} we shall define initial conditions either by
setting $\mu(z)$ to a specified function, or
by requiring the solution $\mu(z)$ to satisfy certain additional
assumptions.

In describing the equilibrium,
we have introduced a natural set
of units, namely $1/\Omega_K$ for time, $c_s$ for velocity, and
$\rho_{g0}$ for density.
For future reference, we note that in the dust-free case, the
solution of equation (\ref{rhog_eq}) is a gaussian with
gas column density $\Sigma_g = \sqrt{2 \pi} \rho_{g0} H_g$,
where $H_g = c_s / \Omega_K$.
In the MSN model, the gas disk aspect ratio is

\begin{equation}
  \frac{H_g}{R} = 4.5 \times 10^{-2}
     \left(\frac{R}{1 \au}\right)^{1/4}
\end{equation}


\section{STABILITY ANALYSIS FOR A THREE-LAYER MODEL}
\label{threelayer_sec}

\subsection{Perturbation Equations and General Solution}
\label{discrete_sec}

Consider a perturbation to the above equilibrium with $
\rho       = \rho_0 + \rho_1,
\vect{v} = \vect{v}_0 + \vect{v}_1,$ and $
P          = P_0 + P_1$.
We assume that the flow is incompressible
($\nabla \cdot \vect{v}_1=0$),
linearize equations (\ref{hd_eq1})-(\ref{hd_eq4}), and, for
simplicity in this first study, we drop terms in
$\partial / \partial x$.
Then, taking $\rho_1,P_1,v_{1x},v_{1y},v_{1z}
\propto \exp[\mi(k_y y -\omega t)]$, the resulting equations can be
combined to obtain a governing equation for $v_{1z}$:

\begin{eqnarray}
  \left[ 1 - \frac{4 \Omega_F^2}{(\omega - k_y v_{0y})^2} \right]
  & \partial_z^2 v_{1z} &
  \nonumber \\
+ \left\{ \partial_z \ln \rho_0
  \left[ 1 - \frac{4 \Omega_F^2}{(\omega - k_y v_{0y})^2} \right]
  - \frac{4 \Omega_F^2 k_y}{(\omega-k_y v_{0y})^3}
  \partial_z v_{0y} \right\}
  & \partial_z v_{1z} &
  \nonumber \\
+ \left[
  \frac{k_y}{\omega-k_y v_{0y}} \partial_z \ln\rho_0 ~\partial_z v_{0y}
  +\frac{k_y}{\omega-k_y v_{0y}} \partial_z^2 v_{0y}
  - k_y^2
  -\frac{k_y^2 g(z) \partial_z \ln\rho_0}
        {(\omega-k_y v_{0y})^2} \right]
  & v_{1z} & = 0,
\label{v1z_eq}
\end{eqnarray}

\noindent
where $\partial_z = \partial / \partial z$.

The usual step at this point is to invoke the Boussinesque
approximation, dropping $\partial_z \rho_0$
everywhere but in the bouyancy term.
Because shear and density gradients are directly associated for the
problem at hand, however,
we shall not make this approximation in this work
(see also discussion in Garaud \& Lin 2004).

\begin{figure}[t]
\epsscale{0.5}
\plotone{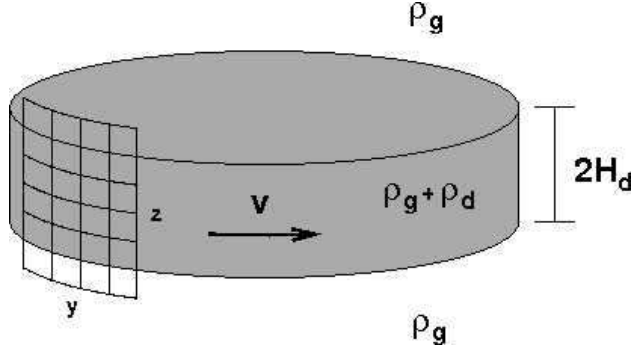}
\parbox{5.5in}{
\caption{
\small
Schematic of the discrete three layer model.
A central gas-dust layer, with density $\rho_g + \rho_d$ and
thickness $2 H_d$,
is sandwiched between two semi-infinite gaseous atmosphere layers
with densities $\rho_g$.
In the reference frame rotating with the atmosphere at $\Omega_F$,
the central layer has velocity $V$ at radius $R_0$.
We consider a small azimuthal-vertical section of the disk, as
shown.
\bigskip
\hrule
\label{three_layer_fig}
}}
\end{figure}

Consider now the three layer distribution shown in Figure
\ref{three_layer_fig}.
The middle layer is composed of the dust and gas mixture, while
the top and bottom layers are dust-free.
We take the initial dust density $\rho_d$ as constant in the
middle layer, and the gas density $\rho_g$ constant across all
layers
(provided $H_d$ is small compared to the total disk thickness $H_g$,
the latter is true more generally).
This three-layer model is a fair approximation to a more realistic
continuous distribution for wavelengths larger than the vertical
velocity and density gradient scales.

In the chosen reference frame, the top and bottom layers are at rest,
while the central layer has a constant velocity
$V = v_{0,max} \rho_d / ( \rho_g+\rho_d) = v_{0,max} \mu / (1+\mu)$.
Since $\partial_z \rho_0 = \partial_z v_{0y}=0$ within each
layer, equation (\ref{v1z_eq}) simplifies to

\begin{equation}
 \partial_z^2 v_{1z} =
   \begin{cases}
      {\displaystyle
      \frac{k_y^2}{1- 4 \Omega_F^2/\omega^2} \,v_{1z}
      }
        & |z| > H_d, \\
      {\displaystyle
      \frac{k_y^2}{1- 4 \Omega_F^2/(\omega - k_y V)^2} \,v_{1z}
      }
        & |z| < H_d.
   \end{cases}
\end{equation}

\noindent
The solution is

\begin{equation}
v_{1z} =
  \begin{cases}
     A_1 \me^{-K_1 z} & z > H_d, \\
     A_{2+} \me^{K_2 z} + A_{2-} \me^{-K_2 z}
       & |z| < H_d, \\
     A_3 \me^{ K_1 z} & z < -H_d,
  \end{cases}
\label{v1zsol_eq}
\end{equation}

\noindent
where

\begin{eqnarray}
K_1 &=& \frac{k_y}
  {\sqrt{ 1- 4 \Omega_F^2 / \omega^2 } }
  \label{k1_eq} \\
K_2 &=& \frac{k_y}
  {\sqrt{ 1- 4 \Omega_F^2 / (\omega- k_y V)^2 } },
  \label{k2_eq}
\end{eqnarray}

\noindent
and where we require that the sign of the square root in equation
(\ref{k1_eq}) is to be chosen such that its real part $\Re(K_1) > 0$.
In order to obtain equation (\ref{v1zsol_eq}), we apply the
boundary condition $v_{1z} \rightarrow 0$ when
$|z| \rightarrow \infty$.
In general, solutions that are either midplane-symmetric or
midplane-antisymmetric are possible.
Some previous work \citep[for example]{ish02}, has explored only the
case in which $v_{1z}$ is an odd function of $z$, so $v_{1z}(0)=0$.
We shall focus much of our attention on the case in which $v_{1z}$
is an even function, and explore the odd case only for reference.

\subsection{Solutions With $v_{1z}$ Even in $z$}
\label{v1zeven_sec}

With $v_{1z}(z)=v_{1z}(-z)$, $A_1=A_3$ and $A_{2+}=A_{2-}=A_2$.
To relate $A_1$ and $A_2$, we use the fact that
the vertical displacement of a lagrangian point,
$\delta z = \mi v_{1z}/(\omega - k_y V)$,
must be a continuous function of $z$
across any interface in the flow.
Applying this condition at the layer discontinuity,
this implies

\begin{equation}
\me^{-K_1 H_d} A_1 = \frac{\omega}{\omega-k_y V}
  A_2 ( \me^{K_2 H_d} + \me^{-K_2 H_d}).
\label{a1a2_eq}
\end{equation}

By integrating equation (\ref{v1z_eq}) across the $z=H_d$ interface,
and substituting the solutions (\ref{v1zsol_eq}), we obtain the
dispersion relation

\begin{equation}
(1+\mu) (\omega-k_y V)^2
  \left[ 1 - \frac{4 \Omega_F^2}{(\omega-k_y V)^2} \right]^{1/2}
  \tanh(K_2 H_d)
+ \omega^2 \left( 1- \frac{4 \Omega_F^2}{\omega^2} \right)^{1/2}
= k_y g \mu.
\label{disprel_eq}
\end{equation}

\noindent
Here, $g = g(H_d)= \Omega_K^2 H_d$.

If we temporarily
disregard the Coriolis terms (setting $\Omega_F=0$;
we shall discuss the case with $\Omega_F \neq 0$ in \S
\ref{chandra_sec}),
then $K_1 = K_2 = k_y$, and we can analytically solve equation
(\ref{disprel_eq}) for $\omega$.
The condition for instability [that the imaginary part of $\omega$,
$\Im(\omega)$, is non-zero] is

\begin{equation}
k_y V^2 > \frac{g \mu}{1+\mu}
  \left[ \frac{1}{\tanh(k_y H_d)} + 1+\mu \right];
\label{insta_eq}
\end{equation}

\noindent
(see also Drazin \& Reid 1981, p. 29).
This can be written in alternate form as

\begin{equation}
k_y H_d > 
  \frac{1+\mu}{\mu}
  \left(\frac{H_d}{H_g}\right)^2
  \left(\frac{c_s}{v_{0,max}}\right)^2
  \left[\frac{1}{\tanh(k_y H_d)} + 1+\mu \right].
\label{instb_eq}
\end{equation}

\noindent
We now define an effective Richardson number for the
discontinuous distribution in the three-layer model,

\begin{eqnarray}
\Ri_{\eff}  &\equiv& \frac{-g ~(\Delta \rho/\Delta z)}
                          {\rho ~(V/\Delta z)^2}
  \nonumber \\
  &=& \frac{1+\mu}{\mu}
      \left( \frac{H_d}{H_g} \right)^2
      \left( \frac{c_s}{v_{0,max}} \right)^2 \nonumber \\
  &=& \frac{\pi}{2}\,\frac{1+\mu}{\mu^3}
      \left( \frac{\Sigma_d}{\Sigma_g} \right)^2
      \left( \frac{c_s}{v_{0,max}} \right)^2,
\label{rich_eq}
\end{eqnarray}

\noindent
where
$\Sigma_d = 2 \rho_d H_d$ is the dust column density, so that

\begin{equation}
\mu = \sqrt{\frac{\pi}{2}} \, \frac{H_g}{H_d} \,
  \frac{\Sigma_d}{\Sigma_g}.
\end{equation}

\noindent
With this definition, the criterion for instability may be written

\begin{equation}
\Ri_{\eff} <
  \frac{k_y H_d \tanh(k_y H_d)}{1 + (1+\mu) \tanh(k_y H_d)}
\label{fullins_eq}
\end{equation}

\noindent
Notice that, for $\mu \gg 1$ and a moderate $k_y H_d$ value,
equation (\ref{fullins_eq}) reduces to $\Ri_{\eff} < (k_y H_d)/\mu$,
i.~e. the layer will still remain stable to wavelengths comparable
to $H_d$ even when $\Ri_{\eff}$ is quite small, for sufficient dust
concentration.


\subsection{Solutions With $v_{1z}$ Odd in $z$}
\label{v1zodd_sec}

In the case with $v_{1z}(z) = - v_{1z}(z)$,
$A_1 = - A_3$ and $A_{2+} = - A_{2-} = A_2$,
and equation (\ref{a1a2_eq}) becomes

\begin{equation}
\me^{-K_1 H_d} A_1 = \frac{\omega}{\omega-k_y V}
  A_2 ( \me^{K_2 H_d} - \me^{-K_2 H_d}).
\end{equation}

\noindent
The resulting dispersion relation for the odd-symmetry mode is

\begin{equation}
(1+\mu) (\omega-k_y V)^2
  \left[ 1 - \frac{4 \Omega_F^2}{(\omega-k_y V)^2} \right]^{1/2}
  \coth(K_2 H_d)
+ \omega^2 \left( 1- \frac{4 \Omega_F^2}{\omega^2} \right)^{1/2}
= k_y g \mu.
\label{disprelodd_eq}
\end{equation}

\noindent
Again, if we consider the $\Omega_F = 0$ case
we can obtain criteria for instability.
These are the same as equations (\ref{insta_eq}), (\ref{instb_eq}),
and (\ref{fullins_eq}) with $\coth(k_y H_d)$ substituted for
$\tanh(k_y H_d)$.
Notice that both the even and odd modes have the same instability
condition for short wavelengths ($k_y H_d \gg 1$),

\begin{equation}
\Ri_{\eff} (2+\mu) < k_y H_d \qquad\mbox{or}\qquad
\frac{g \mu (2+\mu)}{1+\mu} < k_y V^2,
\label{fullinsodd_eq}
\end{equation}

\noindent
which is the same as the
standard criterion \citep[eq. XI.34, for example]{cha61}
for KHI at an interface between two semi-infinite layers.


\subsection{Comparing Even and Odd Modes With $\Omega_F=0$.}
\label{comparison_sec}

\begin{figure}[t]
\epsscale{0.5}
\plotone{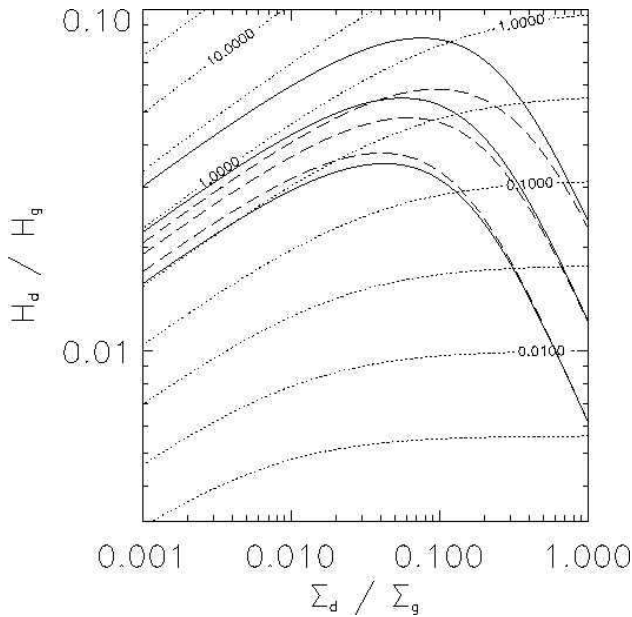}
\parbox{5.5in}{
\caption{
\small
Minimum stable dust layer thickness for global modes with
$\Omega_F=0$ (i.~e. neglecting Coriolis forces),
as a function of the dust-to-gas surface density ratio.
Both even (solid line) and odd modes (dashed line) are shown for
fixed values of $k_y H_d = \pi, \pi/2$ and $\pi/4$ (top, middle and
bottom, respectively).
Curves of constant effective Richardson number (dotted line)
are overlayed.
$\Ri_{\eff} = const.$ is a good stability proxy at low $\Sigma_d /
\Sigma_g$.
\bigskip
\hrule
\label{sdh_fig}
}}
\end{figure}

In order to compare with more realistic models that have a continuous
distribution of dust, we focus on wavelengths that are
comparable to the layer thickness; shorter wavelengths will
not lead to large scale mixing of the dust layer, and longer
wavelengths have smaller growth rates.
Figure \ref{sdh_fig} shows the minimum
stable layer thickness for three different values of $k_y H_d$, for
a range of dust-to-gas surface density ratios
and $v_{0,max}/c_s = 0.1$.
Both even- and odd-symmetry solutions are shown.
In a real system, the dust will (at least initially) likely settle
toward the midplane
at a higher rate than the surface density ratio changes.
So, we can imagine the disk evolving down along a vertical line in
Figure \ref{sdh_fig}, until it reaches the stability edge at a
relevant wavelength.
On a longer time scale, the disk may then move along that edge
towards larger $\Sigma_d / \Sigma_g$ values as gas is
lost to photoevaporation or the dust component is increased due to
radial drift, for example.

The slopes of the critical curves at fixed $k_y H_d$ can be
understood physically.
At small $\Sigma_d / \Sigma_g$, an increase in $\Sigma_d / \Sigma_g$
increases $\mu$, thus increasing the velocity shear since
$V = v_{0,max} \mu / (1+\mu)$.
This forces an increase in $H_d / H_g$ in order to keep the layer
marginally stable;
as a consequence, from equation (\ref{instb_eq}), the critical
$(H_d/H_g) \propto (\Sigma_d / \Sigma_g)^{1/3}$ at small $\mu$.
For large $\Sigma_d / \Sigma_g$, $\mu \gg 1$ and the velocity
difference saturates to $v_{0,max}$;
in this case, the dust layer must get thinner to remain marginally
stable with increasing dust abundance.
At large $\mu$, the critical curve therefore follows
$(H_d / H_g) \propto (\Sigma_d / \Sigma_g)^{-1}$.
These small- and large-$\mu$ scalings of $(H_d/H_g)_{crit}$ with
$\Sigma_d / \Sigma_g$ are the same as have been identified by
\citet{gar04} in their equations (C3) and (C9), respectively.

Lines of constant $\Ri_{\eff}$ are also plotted in Figure \ref{sdh_fig}.
Recall that for a continuous dust distribution, $\Ri < 1/4$ is a
necessary (but not sufficient) condition for the development of the
KH instability.
For our discrete three-layer distribution, $\Ri_{\eff} = const.$ is a good
proxy for the
stability limit of both even and odd modes at fixed $k_y H_d$, but
only at low $\Sigma_d / \Sigma_g$.
For $k_y H_d = \pi/4$, we find that the critical $\Ri_{\eff} \approx
0.3$ at $\Sigma_d / \Sigma_g \lesssim 0.01$, close to the $\Ri =
1/4$ result of \citet{mil61} and \citet{how61}.
At larger $\Sigma_d / \Sigma_g$, the layer is
too heavy to be easily disturbed by a global mode.
In this case, with $\mu \gg 1$, the stability limit
(for either even- or odd-symmetry modes)
is $\Ri_{\eff} = k_y H_d / \mu$.
So, at given $\Ri_{\eff}$ and $H_d/H_g$,
the increasingly short wavelengths that are required for
instability with increasing $\Sigma_d / \Sigma_g$ would make
the disturbance only a surface phenomenon.
Global instabilities only become possible when $H_d / H_g$, and
hence $\Ri_{\eff}$, decreases.
\citet{gar04} find, similarly, that the critical value of $\Ri$
decreases as $\Sigma_d / \Sigma_g$ increases above $\approx 0.01$,
for the value of $v_{0,max}/c_s$ we adopt.

\begin{figure}[t]
\epsscale{0.5}
\plotone{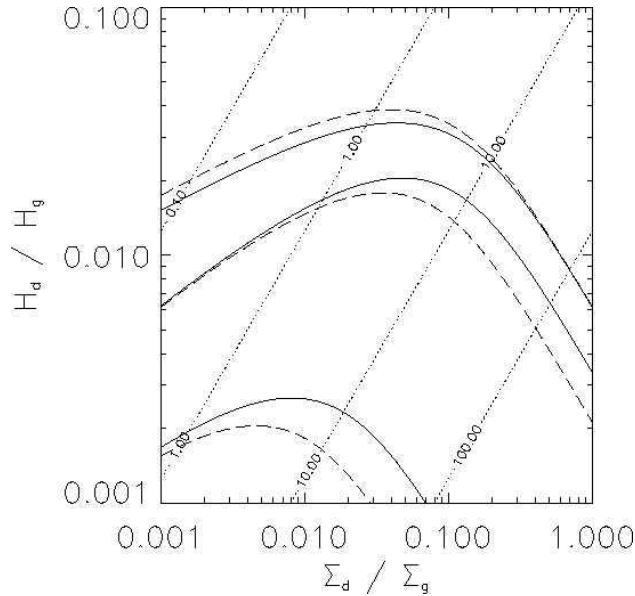}
\parbox{5.5in}{
\caption{
\small
Instability growth rates for $k_y H_d = \pi/4$ and $\Omega_F=0$.
Lines show the even (solid) and odd (dashed) modes
for values of the growth rate (down from the top)
$\omega_I/\Omega_K = 0$ (stability edge), 1 and 10.
Dotted lines follow $\mu = const$.
\bigskip
\hrule
\label{shgrowth_fig}
}}
\end{figure}

For large $\Sigma_d / \Sigma_g$ and fixed $k_y H_d$,
Figure \ref{sdh_fig} shows that the maximum unstable $H_d / H_g$
has the same wavelength for both the even and odd modes.
But the growth rates are not the same, as shown in
Figure \ref{shgrowth_fig}.
As the layer thickness decreases at constant surface density, even
modes at a given $k_y H_d$ will develop more rapidly than the
corresponding odd modes.
Or, conversely, an even mode with longer wavelength (and more
efficient dilution of the dust layer) will become excited at the
same time as a (less-efficient) odd mode with shorter wavelength.
This can be expected, since the incompressibility condition implies,
for the odd mode, that a horizontal converging flow must be
generated in the midplane at the position where the interface
displacement makes the layer thicker ($\delta z/z > 0$).
Since the even mode does not need to channel energy into this
requirement, it can grow faster.
On the other hand, since the odd mode does not need to vertically
displace the whole layer, it can remain unstable for larger
$H_d / H_g$ values, at which gravity is stronger at the layer
interface.


\subsection{Even Mode, Including Coriolis Forces ($\Omega_F \neq 0$)}
\label{chandra_sec}

In \S \ref{v1zeven_sec} and \S \ref{v1zodd_sec} we studied the
solutions of equations (\ref{disprel_eq}) and (\ref{disprelodd_eq})
obtained when the Coriolis forces are neglected.
While the solutions presented were for the exact equations, we have
found that taking the long-wavelength limit $k_y H_d \ll 1$
yields a good approximation to the stability even at values of $k_y
H_d$ relatively large ($\sim 10$).
When the Coriolis terms are included ($\Omega_F \neq 0$), the
dispersion relation retains its simple character when $k_y H_d \ll
1$, so for this section we concentrate on that limit.%
\footnote{For short wavelengths, $k_y H_d \gg 1 \Rightarrow
\tanh(K_2 H_d) \approx 1$, and
equation (\ref{disprel_eq}) yields
the same dispersion relation as the two layer problem
with rotation studied by \citet[\S 105]{cha61};
see also \citet{hup68}.
}

\begin{figure}[t]
\epsscale{0.5}
\plotone{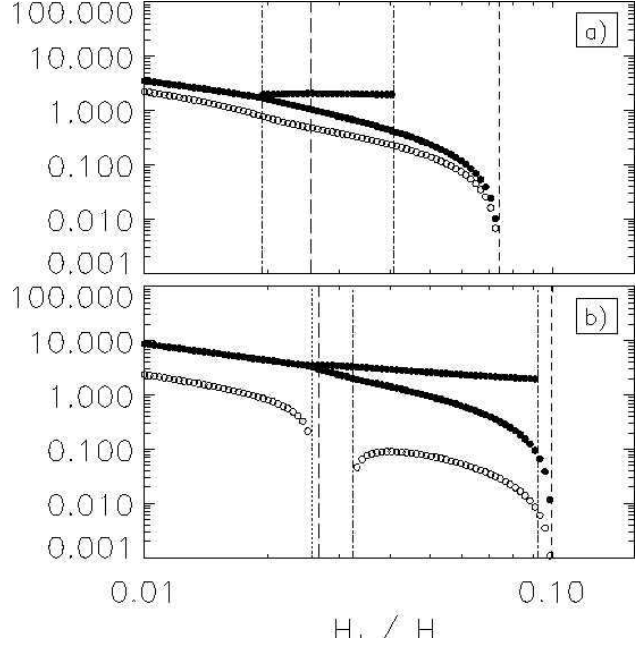}
\parbox{5.5in}{
\caption{
\small
Roots of equation (\ref{disprel_long_eq}), as a function of
$H_d / H_g$, for $k_y H_d = \pi/3$ with $\mu = 1$ ({\it a})
and $\mu=10$ ({\it b}).
Filled circles show $\Re(\omega)$, and open circles show
$\Im(\omega)$.
Vertical lines mark critical $H_d / H_g$ values
(see Appendix).
There is no solution for $\Ri_{\eff} > (k_y H_d)^2$, i.~e. large
$H_d / H_g$.
\bigskip
\hrule
\label{chandra_fig}
}}
\end{figure}

In the long wavelength regime,
equation (\ref{disprel_eq}) can be written as

\begin{equation}
(1+\mu) (\omega-k_y V)^2
  k_y H_d
+ \omega^2 \left( 1- \frac{4 \Omega_F^2}{\omega^2} \right)^{1/2}
= k_y g \mu,
\label{disprel_long_eq}
\end{equation}

\noindent
where we require the real part of the square root to be
positive.
The procedure used to solve
equation (\ref{disprel_long_eq}) is presented in the Appendix, and
the corresponding roots are shown in Figure \ref{chandra_fig}.
The number and nature of the roots can change only at critical
$H_d / H_g$ values (as described in the Appendix), marked in Figure
\ref{chandra_fig} by the vertical lines.

For the parameters presented in Figure \ref{chandra_fig}b,
there is a range of values of $H_d / H_g$ with only real
solutions, which implies stability.
Nevertheless, we consider the existence of this gap
of limited practical significance since the gap closes for
longer wavelengths (which yield more efficient vertical mixing).
More significant is the fact that
there are no solutions, real or complex, when
$(k_y V/\Omega_K)^2 < \mu/(1+\mu)$, or
$k_y v_{0,max} / \Omega_K < (1+1/\mu)^{1/2}$
(right hand side of Figure \ref{chandra_fig}).
The transition critical point corresponds to $\omega/\Omega_K = 0$.
For given (order unity) $k_y H_d$, and any $\Sigma_d / \Sigma_g$,
sufficiently small $\mu$ guarantees that this condition holds.
Thus, using equation (\ref{rich_eq}),
a necessary (but not sufficient) condition for instability at
long wavelengths appears to be

\begin{equation}
\Ri_{\eff} < (k_y H_d)^2.
\label{rotins_eq}
\end{equation}

\noindent
Comparison of equation (\ref{rotins_eq}) and its equivalent for the
$\Omega_F=0$ case (eq. [\ref{fullins_eq}]) shows that a larger
$\Ri_{\eff}$ is required for stability when the Coriolis terms are
turned on: for $k_y H_d = \pi/3$, the critical value
for $\Ri_{\eff}$ is 1.1 when $\Omega_F \neq 0$, compared to 0.32,
0.14, and 0.09 for $\mu=1, 5,$ and 10, respectively
when $\Omega_F  =  0$.
Notice that both conditions
reduce to the same criterion for instability when $k_y H_d$ and
$\mu \ll 1$.
However, small $\mu$ and $\Ri_{\eff} < (k_y H_d)^2 < 1$ are satisfied
only for $\Sigma_d / \Sigma_g \ll 0.01$, i.~e., for cases with
much lower metallicity than that expected in the proto-solar nebula.


An interesting feature of equation (\ref{rotins_eq}) is that
(unlike eq. [\ref{fullins_eq}])
it does not explicitly depend on the dust abundance, $\mu$.
This is due to the fact that, at the onset of instability
for the $\Omega_F \neq 0$ case,
the phase velocity vanishes [$\Re(\omega) \rightarrow 0$].
This has two consequences.
First, a larger fraction of the heavier layer's momentum is
available to help destabilize the flow (and so, the layer remains
unstable for larger $\Ri_{\eff}$).
And second, a vanishingly small $\omega$ means a small $K_1$
(eq. [\ref{k1_eq}]), so that vertical pressure gradients in the gas
layer are very small.
So, as the dust/gas mixture flows through the sinuous focus of
the perturbed midplane layer, only the
centrifugal force in the vertical direction
($V^2 k_y^2 \, \delta z$) and gravity
($\Omega_K^2 \, \delta z$) are involved in the force balacing, while
gas pressure gradients are neglegible.
In contrast, for the $\Omega_F = 0$ case, $\Re(\omega)$ is
non-vanishing at the onset of instability, yielding a finite $K_1$
value.
This implies non-neglegible vertical pressure gradients,
which introduce a
dependence on the dust-to-gas density ratio in the force balance.
Small or vanishing $\Re(\omega)$ is possible only in the $\Omega_F
\neq 0$ case, when the Coriolis forces (from $v_{1x}$)
are able to partly of fully compensate for azimuthal pressure forces.

We can alternatively write the instability condition for
$\Omega_F \neq 0$ in terms of more basic
physical parameters by considering equation (\ref{rich_eq}) in the
large and small $\mu$ limits:

\begin{equation}
\frac{H_d}{H_g} <
  \begin{cases}
    {\displaystyle
      \left(\frac{\pi}{2}\right)^{1/6}
      \left(\frac{\Sigma_d}{\Sigma_g}\right)^{1/3}
      \left(\frac{v_{0,max}}{c_s}\right)^{2/3} (k_y H_d)^{2/3}
    }
    & \mbox{for } \mu \ll 1, \\
    {\displaystyle
      \frac{v_{0,max}}{c_s} \, k_y H_d
    }
    & \mbox{for } \mu \gg 1.
  \end{cases}
\label{rotinsb_eq}
\end{equation}

\noindent
Notice that the minimum stable layer thickness approaches a constant
value as $\Sigma_d / \Sigma_g$ increases, as shown by the dotted
lines of Figure \ref{sdh_fig}.
This stands in contrast with the instability criterion for the
$\Omega_F = 0$ case (see equation [\ref{instb_eq}]
and Figure \ref{sdh_fig}),
for which $(H_d/H_g)_{crit} \propto (\Sigma_d / \Sigma_g)^{-1}$ as
the surface density increases.
Thus, for large $\mu$, the instability criterion is $\Omega_K <
v_{0,max} k_y$ for $\Omega_F \neq 0$, compared to
$(\mu k_y H_d)^{1/2} \Omega_K < v_{0,max} k_y$ for $\Omega_F = 0$.

While we have concentrated on the long-wavelength limit in the
discussion of this section, we note that the general dispersion
relation of equation (\ref{disprel_eq}) can easily be solved in the
limit $|\omega| \rightarrow 0$.
When $\Omega_F \neq 0$, an unstable solution will appear when
$\Ri_{\eff}$ passes below the value

\begin{equation}
\Ri_{crit} = (k_y H_d)^2
  \tanh \left[ \frac{k_y H_d}{\sqrt{1-4\Omega_F^2/(k_y V)^2}} \right]
  \frac{\sqrt{1 - 4\Omega_F^2/(k_y V)^2}}{k_y H_d},
\end{equation}

\noindent
with growth rate $\propto (\Ri_{crit}-\Ri_{\eff})\Omega_F$.
Thus, the necessary criterion for instability in the long-wavelength
limit is quite similar to a more general {\it necessary and
sufficient} criterion for a marginally-unstable mode to appear.
Since $\tanh(\chi)/\chi > 1$, $\Ri_{crit}$ is in practice slightly larger
than the value deduced in the long-wavelength limit.

We note that for the odd-symmetry case, the critical $\Ri_{\eff}$ is
given by the same expression as above, with $\coth(x)$ substituted for
$\tanh(x)$; for long wavelengths, the instability criterion is
therefore
$(k_y V)^2 > 4 \Omega_F^2 + k_y^2 g H_d \mu/(1+\mu)$, or, using
$\Omega_F \approx \Omega_K$,

\begin{equation}
\Ri_{\eff} < \frac{(k_y H_d)^2 \mu}{4(1+\mu) + (k_y H_d)^2 \mu}.
\end{equation}

\noindent
This is always smaller than than the criterion for the even-symmetry
case, and $\Ri_{crit} \rightarrow (k_y H_d)^2/[4+(k_y H_d)^2]$ for
$\mu \gg 1$.

Finally, we note that the conclusions based on the odd-symmetry
three-layer system are similar to those for the two-layer system
considered by \citet{cha61}.
For the two-layer system with rotation, instability first becomes
possible when

\begin{equation}
(k_y V)^2 > 2 \Omega_F^2
  + \left[ 4 \Omega_F^4 + k_y^2 g^2
    \left(\frac{\mu}{1+\mu}\right)^2\right]^{1/2}.
\end{equation}

\noindent
This can also be written, taking $\Omega_F \approx \Omega_K$, as

\begin{equation}
\Ri_{\eff} < \frac{(k_y H_d)^2 \mu}
  {2(1+\mu) + \sqrt{ 4(1+\mu)^2 + (k_y H_d)^2 \mu^2}}.
\end{equation}

\noindent
For $\mu \gg 1$, this yields
$\Ri_{crit}= (k_y H_d)^2 / [2+\sqrt{4+(k_y H_d)^2}]$,
very similar to the odd-symmetry three-layer case.


\section{NUMERICAL SIMULATIONS}
\label{numerics_sec}

\subsection{Basic Numerical Methods}
\label{setup_sec}

With physical understanding developed via the instability
analysis of the simple three-layer model in \S \ref{threelayer_sec},
we now turn to fully non-linear simulations
with a continuous dust distribution.
Our solutions of equations (\ref{hd_eq1})-(\ref{hd_eq4}) are
obtained using a version of ZEUS \citep{sto92},
a finite difference, time explicit, operator split, eulerian code
for numerical hydrodynamics.
For the purpose of this investigation, we implemented source terms
corresponding to the Coriolis force and
the background radial pressure gradient.
We treat this background radial pressure gradient as constant in
time and space,
$-\partial P_0/\partial R \equiv 2 \Omega_F \rho_{g0} v_{0,max}$.

As described in \S \ref{linear_sec}, in the strong-coupling limit
both the gas and dust separately obey the continuity equation
(eqs. [\ref{hd_eq1}] and [\ref{dust_eq}]).
We implement this requirement in ZEUS by updating $\rho_d$ using
the same transport algorithm as used for the total density,
$\rho = \rho_g+\rho_d$.
We have tested this implementation by advecting
a one dimensional, gaussian-shaped ``dust
cloud'' in pressure equilibrium with its surroundings.
A constant velocity was given to the full grid (200 zones)
which advected the dust along five cloud-lengths,
and across the periodic boundaries.
This experiment was repeated for each coordinate direction of the
code.
In all cases, the total dust mass remained constant
and the rms deviation from the initial gaussian shape was
only 0.4\%.

Since ZEUS was not designed to solve incompressible flow problems,
we need to verify that its algorithms yield acceptable solutions in
the regime of interest, and in particular, can properly evolve
Kelvin-Helmholtz instabilities.
To test this, we simulated a velocity shear slab with a
discrete jump in velocity and density, similar to the setup in
\S \ref{threelayer_sec}, with the Coriolis terms turned off.
These simulations were performed in a two-dimensional $(y-z)$ plane
with initial velocity $\vect{v} = v_y(z) \vect{\hat e}_y$.
The initial conditions were perturbed with an eigenfunction of the
even mode, having a specified wavelength.
We then measured the growth rate of the specified perturbed mode,
while it remained in the linear phase and before other modes
(faster growing with smaller wavelengths, seeded by
grid-size noise), interfered in any
obvious way.
We compared this growth rate with the
analytical value for a variety of perturbation amplitudes,
obtaining an agreement of the order of 10\%.

We now move on to the initialization of our
full non-linear, continuous simulations.
For each model, we set up a disk atmosphere in vertical hydrostatic
equilibrium, and with azimuthal velocity $v_y(z)$
compensating radial gravity and the background pressure gradient.
As described in \S \ref{linear_sec},
setting the value of $\mu(z)=\rho_d(z)/\rho_g(z)$ defines all the
hydrodynamic variables in the equilibrium state.


\subsection{Experiments With Prescribed $\mu(z)$}
\label{mu_sec}

The experiments described in this section are initialized by simply
defining $\mu$ as an arbitrary function of $z$, namely,

\begin{equation}
\mu(z) = \mu_0 \sech^2(z/H_d),
\label{sech_eq}
\end{equation}

\noindent
with $\mu_0 = \mu(0)$.
Three parameters uniquely define a model:
$\mu_0$, $v_{0,max}/c_s$ and $H_d / H_g$
(for consistency with \S \ref{threelayer_sec},
in practice we shall use $\Sigma_d / \Sigma_g$ instead of $\mu_0$).
In these experiments,
the simulation domain is a square grid of size $20 H_d$ 
($257 \times 257$ grid points),
with periodic $x$- and $y$-boundaries, and closed $z$-boundaries.%
\footnote{We also tested open $z$-boundaries, and found no
significant differences.}
The equilibrium state is perturbed
by random velocities with a maximum amplitude of $10^{-3} c_s$.
In all the cases, $v_{0,max} = 0.1 c_s$ which, for
MSN parameters, corresponds to a radial location at $R \approx 3.5 \au$.

\begin{figure}[!p]
\epsscale{0.9}
\plottwo{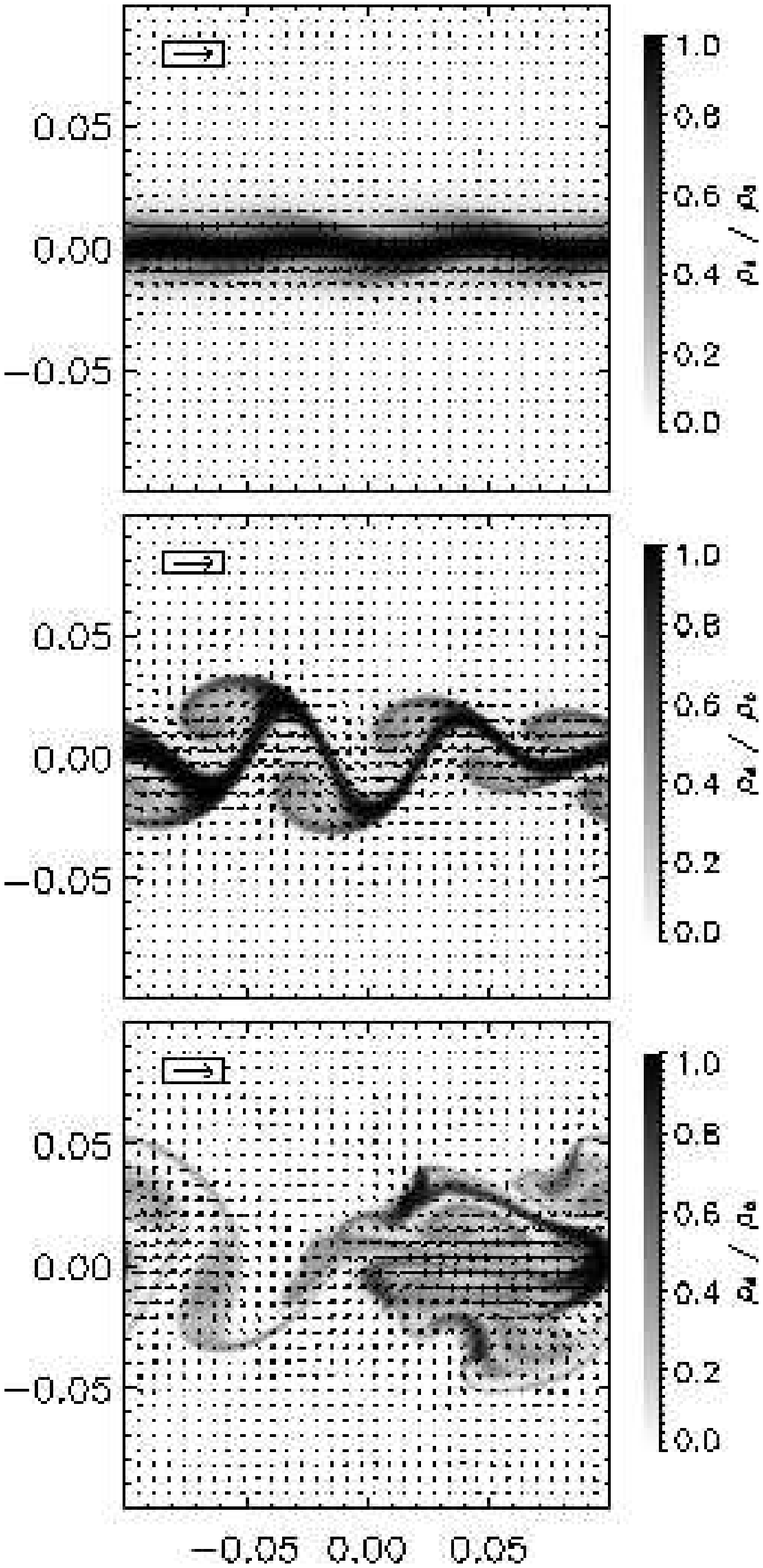}
        {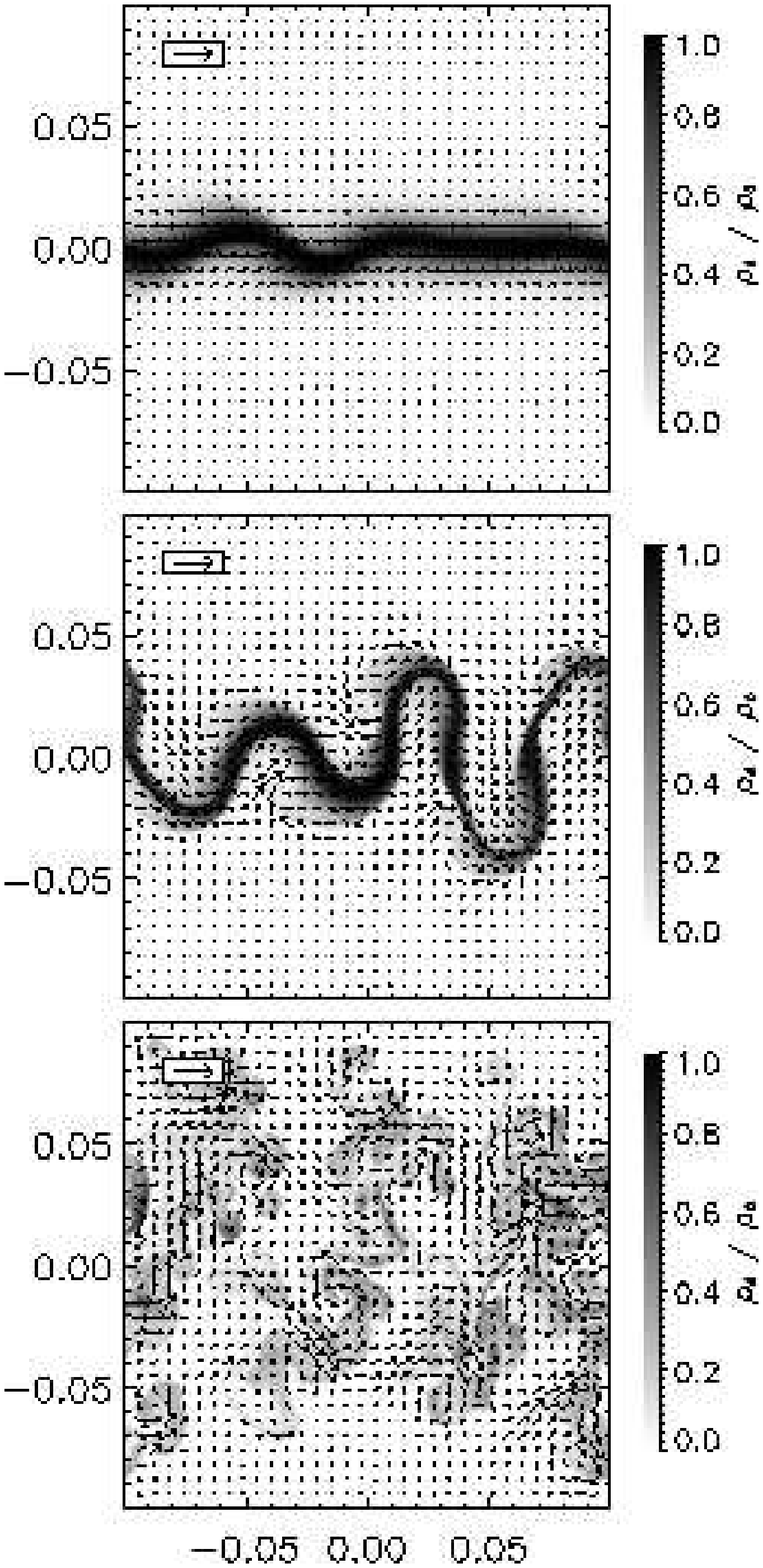}
\parbox{5.5in}{
\caption{
\small
Sample snapshots from a simulation with prescribed $\mu(z)$, both with
Coriolis terms turned off ($\Omega_F=0$, left column) and turned on
($\Omega_F \neq 0$, right column).
The grayscale shows dust density and the arrows show the velocity
in the dust-free reference frame; the arrow in the box 
represents a magnitude of $v_{0,max}/c_s$.
Coordinate axes are in units $H_g$.
Both models have $H_d = 0.01 H_g$, $\Sigma_d = 0.008 \Sigma_g$, and
$v_{0,max} = 0.1 c_s$.
Snapshots correspond to $t = 1.5, 2.2$ and 2.9 $t_{orb}$
for the left column, and $t= 1.8, 2.3$ and 2.8 $t_{orb}$ for the right
column.
\bigskip
\hrule
\label{example_fig}
}}
\end{figure}

Figure \ref{example_fig} shows results from the
simulations with $H_d = 0.01 H_g$ and $\Sigma_d = 0.008 \Sigma_g$,
both with the Coriolis terms turned off (left column;
$\Omega_F=0$) and with the
Coriolis terms turned on (right column; $\Omega_F \neq 0$).
The density snapshots presented are
$t = 1.5, 2.2$ and $2.9 t_{orb}$ ($\Omega_F=0$),
and $t = 1.8, 2.3$ and $2.8 t_{orb}$ ($\Omega_F \neq 0$),
where $t_{orb} = 2 \pi / \Omega_K$ is the orbital time.
The initial linear growth phase is very similar for
both cases, with a sinusoidal displacement of the dust layer
consistent with the even (midplane-crossing) mode of the instability.
Notice that the instability develops three
wavelengths in the computation domain, which corresponds to
$k_y H_d \approx 0.95$.
We also performed simulations with 3 and 5 times larger grids in the $y$
direction (at lower resolution), and found that the same
approximate $k_y H_d$ mode predominated.
The fastest growing mode in the analysis presented in
\citet{ish02,ish03} has a similar $k_y H_d$ value.

In the $\Omega_F=0$ case, once the displacement is large enough,
drag with the dust-poor fluid erodes the tips of the
displaced layer, generating familiar backward-facing
KH rolls.
The flow around the rolls pushes the dust toward the peaks, where it
accumulates until the layer breaks.
After the wave breaks, the dust mixes with the surrounding gas, lowering the
dust abundance and the velocity shear, and generating a more-or-less
homogeneous layer with a thickness similar to the wavelength of the
linear growth.
(For some parameter choices, this layer is in turn unstable,
and again developes a linear growth phase.)
Remnant non-organized motions generate diffusion that further
increase the layer thickness over longer time-scales.

\begin{figure}[t]
\epsscale{1.0}
\plottwo{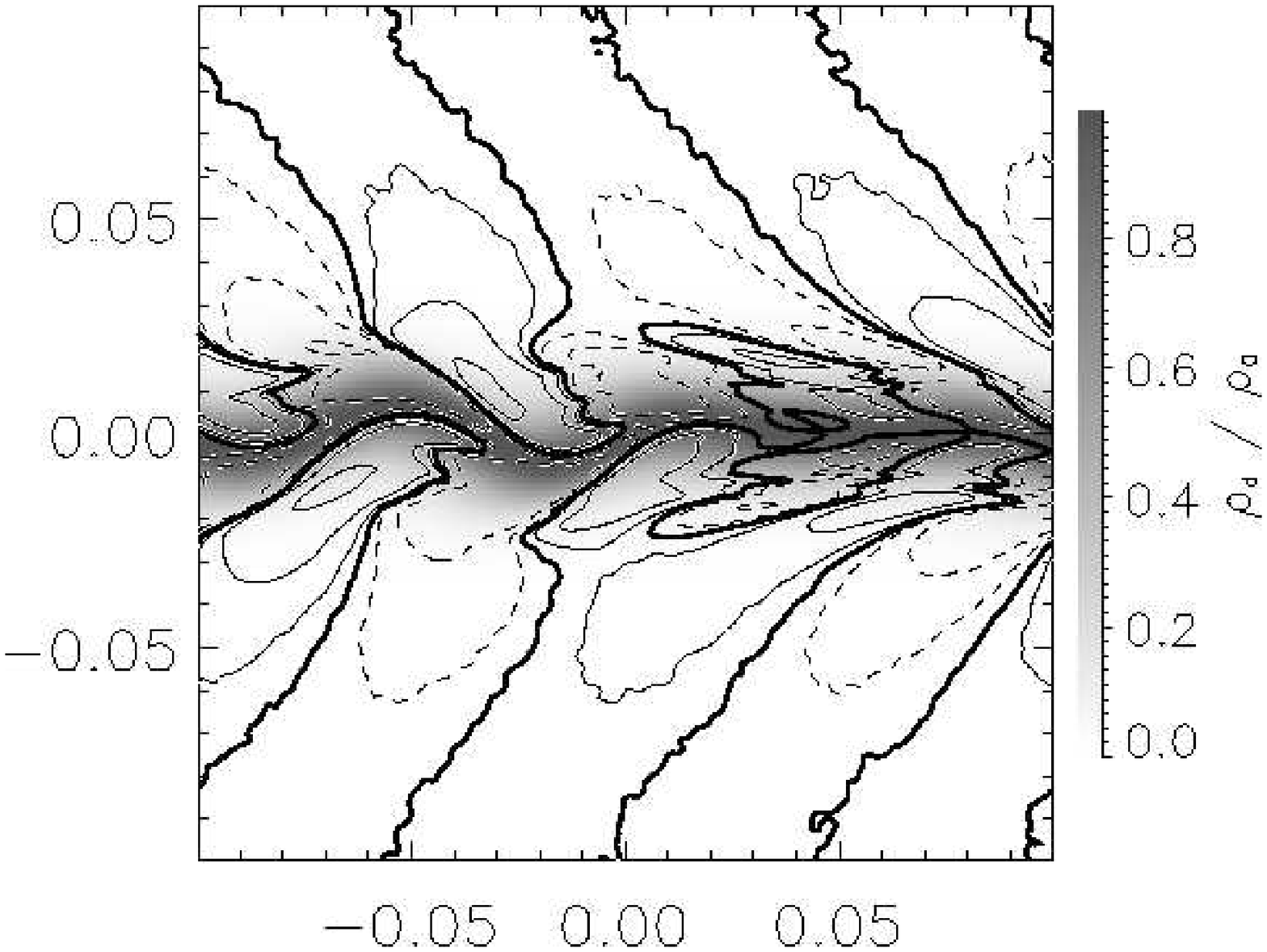}
        {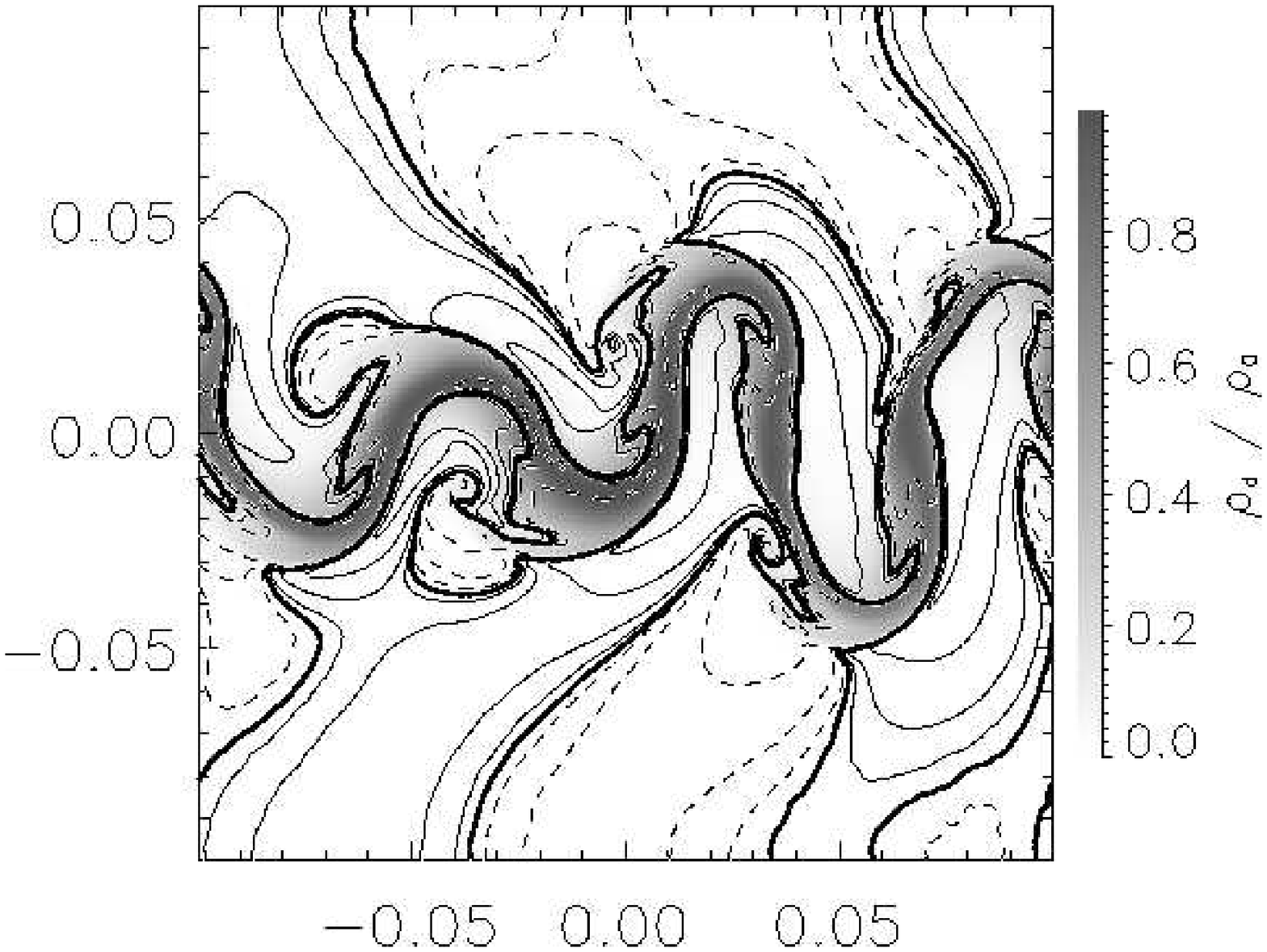}
\parbox{5.5in}{
\caption{
\small
Snapshots from the $\Omega_F \neq 0$ simulation shown in Figure
\ref{example_fig}, at $t=1.8$ (left) and $ 2.3 t_{orb}$ (right).
The grayscale shows dust density, and lines show contours of $v_x$:
thin solid for $v_x>0$, thick solid for $v_x=0$, and dashed for
$v_x < 0$.
As the perturbation grows, the dust layer experiences the slower-moving
high latitude gas as a head-wind.
This gas steals angular momentum from the dust layer, which flows radially
inward, while the dust-poor gas flows outward.
\bigskip
\hrule
\label{radvel_fig}
}}
\end{figure}

In contrast to the $\Omega_F=0$ model, after the linear growth phase
the $\Omega_F \neq 0$ case has a period of very rapid evolution.
Unlike the $\Omega_F = 0$ case, the dust accumulates in the midplane
at the nodes of the displaced layer.
As the vertical displacement grows, the leading faces of the
perturbed dust layer encounters the slower-moving high-latitude gas.
As the gas is pushed by the dust, it gains angular momentum and flows
radially outward (see Figure \ref{radvel_fig}).
Correspondingly, the dusty layer looses angular momentum and
flows radially inward, increasing its azimuthal
velocity.
As a consequence,
the displaced peaks move faster than the
gas in the midplane, generating forward leaning structures.
When the initial density layer looses integrity, it forms a new
layer thicker than the one generated in the $\Omega_F=0$ case.
The subsequent diffusive growth also develops more rapidly.

\begin{figure}[t]
\epsscale{0.5}
\plotone{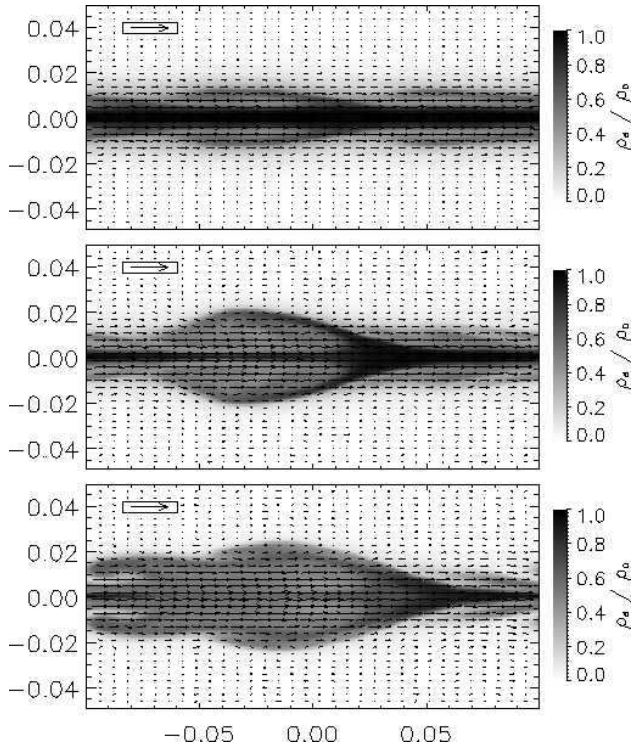}
\parbox{5.5in}{
\caption{
\small
Simulation with prescribed $\mu(z)$ for the $\Omega_F=0$ case,
forcing the odd mode of the
instability by placing a reflecting boundary at $z=0$.
From top to bottom, snapshots show dust density (grayscale) and
velocity field (vectors)
at $t = 2.5, 5.0$ and 7.5 $t_{orb}$.
Notice the smaller vertical extention of the grid, compared
with Figure \ref{example_fig}.
\bigskip
\hrule
\label{example_odd_fig}
}}
\end{figure}

Nearly every set of parameters we tested triggered growth dominated
by an even mode, consistent with expectations from
the discussion in \S \ref{comparison_sec}.
In order to induce the development of the odd mode, we forced the
vertical symmetry by placing a boundary of the simulation at $z=0$,
for the $\Omega_F=0$ case, at the same spatial resolution.
Results for this experiment are shown in Figure \ref{example_odd_fig}
(the grid is doubled in this plot for easier comparison with Figure
\ref{example_fig}).
There are several differences evident between these examples of even
and odd modes.
First, the odd mode developes longer wavelengths, fitting only two
wavelengths in the domain, evolving later to a single wavelength.
Second, despite the longer wavelength,
the extent of the layer after the non-linear growth phase
is thinner, with the dust accumulated at the midplane.
Third, this (and other) odd mode simulations evolve much more slowly
than the even mode ones.
This odd mode simulation has a much larger latency time,
despite the fact that the initial perturbations
are a factor of 10 larger than the even mode example.
The dominant wavelengths that develop for the even modes
have larger growth rates than the ones for the odd modes:
for this case, $0.58 \Omega_K$ vs. $0.05 \Omega_K$
(the procedure to measure these growth rates is described in
\S \ref{rich_sec}).

In addition to the sample cases shown, we have performed a large
number of similar models with both $\Omega_F=0$ and $\Omega_F \neq 0$,
and either full or ``half'' grids.
We have covered a range $H_d/H_g=0.001$ through $0.08$, with a constant
surface density ratio $\Sigma_d / \Sigma_g = 0.008$.
Overall, we find a confirmation of the above results: the layer
evolves more rapidly when $\Omega_F \neq 0$, and more slowly when we
force the development of the odd mode.
We also found that, for $\Omega_F = 0$ and this $\Sigma_d / \Sigma_g$ value,
the layer becomes stable for $H_d/H_g \gtrsim 0.02$.
When $\Omega_F \neq 0$, the layer appeared to be always unstable,
although the evolution times are quite small for the thickest
layers.


\subsection{Experiments With $\Ri = const$}
\label{rich_sec}

The initialization procedure adopted in \S \ref{mu_sec} is quite
simple, but also quite arbitrary in terms of the detailed functional
form of the dust distribution.
As an alternative approach, we can adopt a prescription for the
initial conditions which satisfies additional physically-motivated
constraints.
A natural choice is to set $\Ri$ to a constant value everywhere in
the flow;
this choice is motivated by the common conception that low-$\Ri$
flows are unstable, and high-$\Ri$ flows are stable.
In particular, \citet{sek98} and \citet{you02} have argued that the
disk may evolve to a state where $\Ri=1/4$ (see \S \ref{intro_sec}).

Using the definition of the Richardson number in equation
(\ref{ri_eq}), and equations (\ref{vy0_eq}) and (\ref{rhog_eq}),
we can obtain a differential equation for $\mu$
(for a given constant $\Ri$):

\begin{equation}
\frac{\dif \mu}{\dif z}
  = \frac{- z (1+\mu)^3}{2 \Ri\, H_g^2 (v_{0,max}/c_s)^2}
  \left[ 1 + \sqrt{ 1+ \frac{4 \Ri (v_{0,max}/c_s)^2}{1+\mu}}\right].
\label{riconst_eq}
\end{equation}

\noindent
An approximate solution to this equation can be found
provided that $4 \Ri\, (v_{0,max}/c_s)^2 / (1+\mu)
\ll 1$ (true for the parameter space of interest).
Using this approximation,

\begin{equation}
1+\mu(z) = \left[ \left(\frac{1}{1+\mu_0}\right)^2
         + \frac{ (z/H_g)^2 }{\Ri\, (v_{0,max}/c_s)^2}
  \right]^{-1/2}.
\label{muapprox_eq}
\end{equation}

\noindent
Again, three parameters define a model:
$\mu_0$ (or alternatively, $\Sigma_d / \Sigma_g$),
$v_{0,max}/c_s$, and $\Ri$.
This equation sets a finite extension for the dust layer.
Setting $\mu(z)=0$, we obtain

\begin{equation}
\frac{z_{max}}{H_g}
  = \left\{ \Ri \left[ \frac{v_{0,max}}{c_s} \right]^2
  \left[ 1-\frac{1}{(1+\mu_0)^2}\right]
  \right\}^{1/2}.
\label{zmax_eq}
\end{equation}

We use this setup to study the stability of the dust layer under
varying conditions.
Cases with and without the Coriolis terms were performed.
For our array of simulations, we adopt $v_{0,max}/c_s =0.1$, and
explore a range of initial values of
$\Sigma_d / \Sigma_g$ and $\Ri$.
Our parameter grid covered
$\Sigma_d / \Sigma_g = 0.006$ to $0.178$, and
$\Ri = 0.1$ to $1.0$ for the $\Omega_F=0$ cases,
$\Ri = 1.0$ to $10.0$ for the $\Omega_F \neq 0$ cases.
The domain of these simulations extended
$\pm 10 z_{max}$ in $y$, and $\pm 2 z_{max}$ in $z$
(we did not need a large $z$ extension
since these experiments were designed
to follow only the linear part of the instability,
as opposed to the full evolution of \S \ref{mu_sec}).
Our standard resolution was using $200 \times 40$ grid points,
with higher resolution models tested
for selected cases in order to verify the results.
Each simulation was followed through $t/t_{orb} = 400 z_{max}/H_g$,
which corresponds to values between $9$ and $40 t_{orb}$ for $\Ri$
between 0.1 and 1.0.
The initial equilibrium setup was perturbed with random
velocities with a maximum amplitude of $10^{-3} c_s$.

\begin{figure}[t]
\epsscale{1.0}
\plottwo{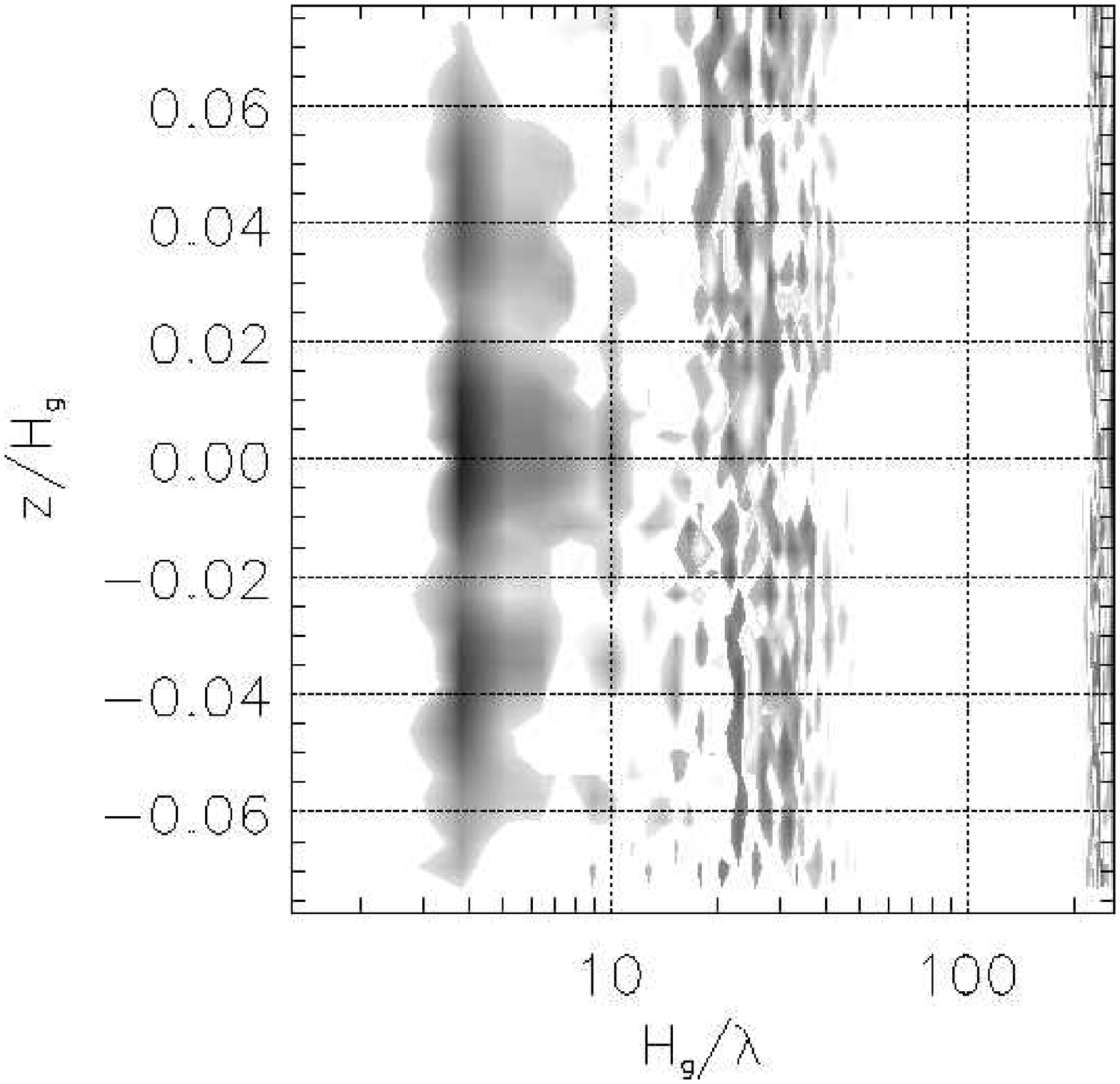}
        {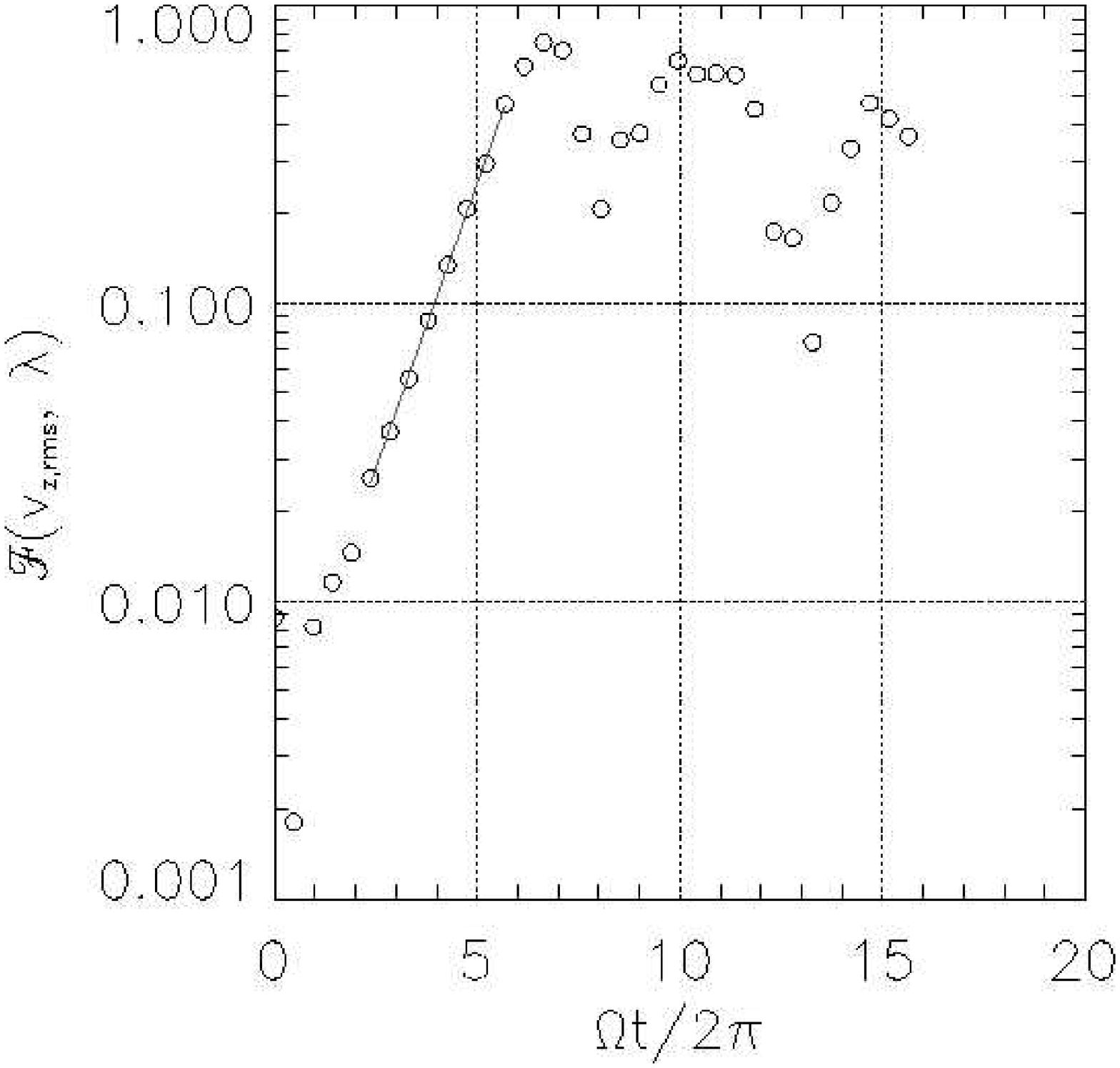}
\parbox{5.5in}{
\caption{
\small
{\it Left}: Fourier transform of $v_z$ along the $y$ direction, as a
function of $z$, for the $\Ri = 0.158$, $\Sigma_d / \Sigma_g=0.056$,
$\Omega_F=0$ case, after 4.27 orbital times.
{\it Right}: Amplitude of the $\lambda = 0.26 H_g$ mode, in
arbitrary units,
as a function of time, for $z=0$.
The linear growth phase of the instability (solid line)
shows a growth rate $\omega_I = 0.14 \Omega_K$.
\bigskip
\hrule
\label{fourier_fig}
}}
\end{figure}

After exploring several possibilities, we found that a practical
way of assessing the stability of the dust layer was by
calculating the Fourier transform of $v_z$ along the $y$ direction,
as a function of $z$.
When the layer is unstable,
the fastest growing mode is easily identified from this plot;
its amplitude can then be examined as a function of time.
In most cases, the linear growth phase of the
instability is clearly evident, and a growth rate can be
measured.
An example is presented in Figure \ref{fourier_fig}.

As expected, the empirically-determined region of stability in the
parameter space does not have a sharp edge, but instead growth
rates decrease until our procedure becomes too insensitive to
measure them.
For practical purposes, we denote
as ``stable'' those simulations that do not show
growth within the allotted run time, or else for which the
measured growth rate is lower than a fiducial value, set to
$\omega_I = 0.05 \Omega_K$;
this growth rate corresponds to a ten-fold
increase in amplitude over 10 orbital times.

\begin{figure}[t]
\epsscale{1.0}
\plottwo{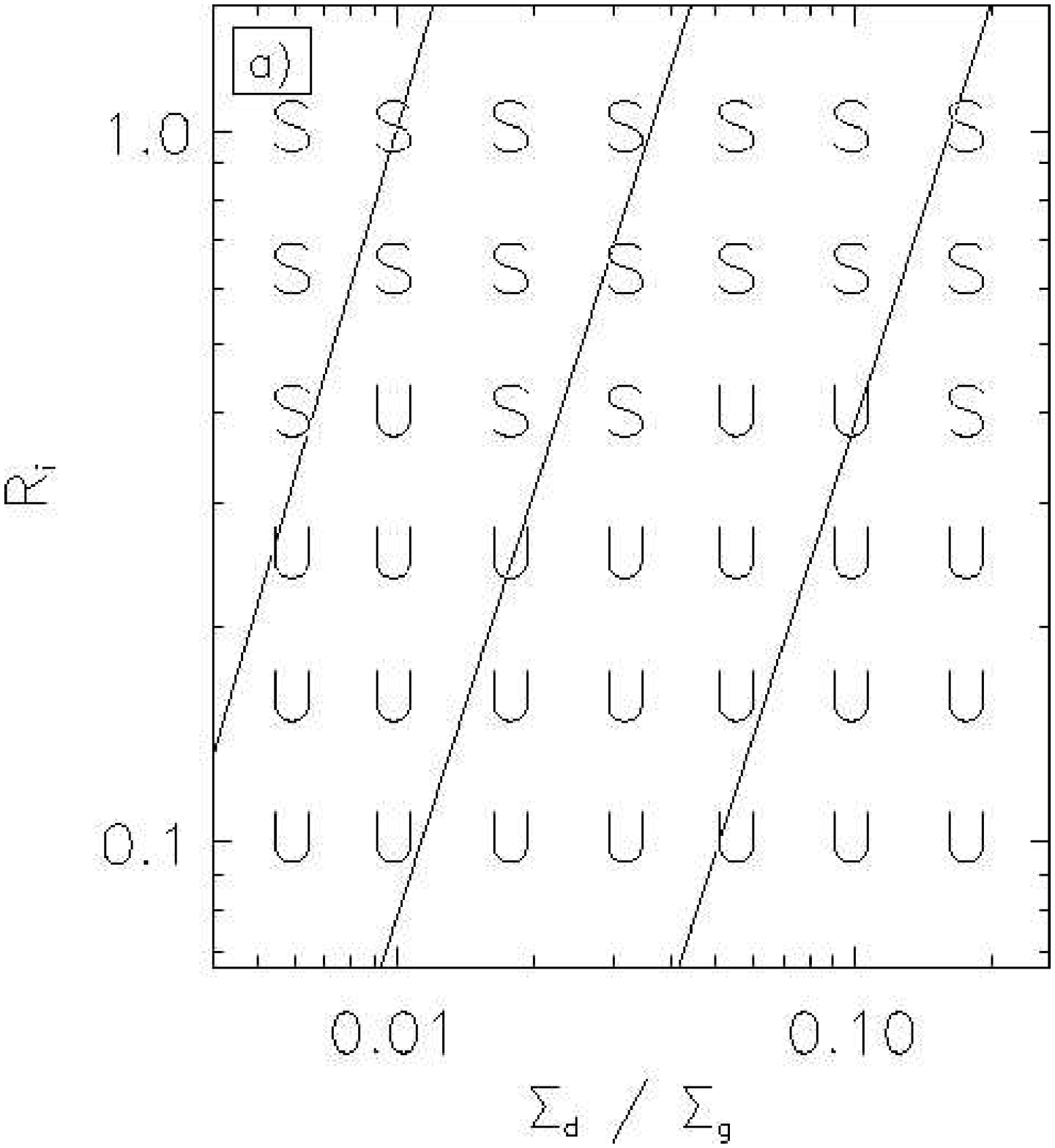}
        {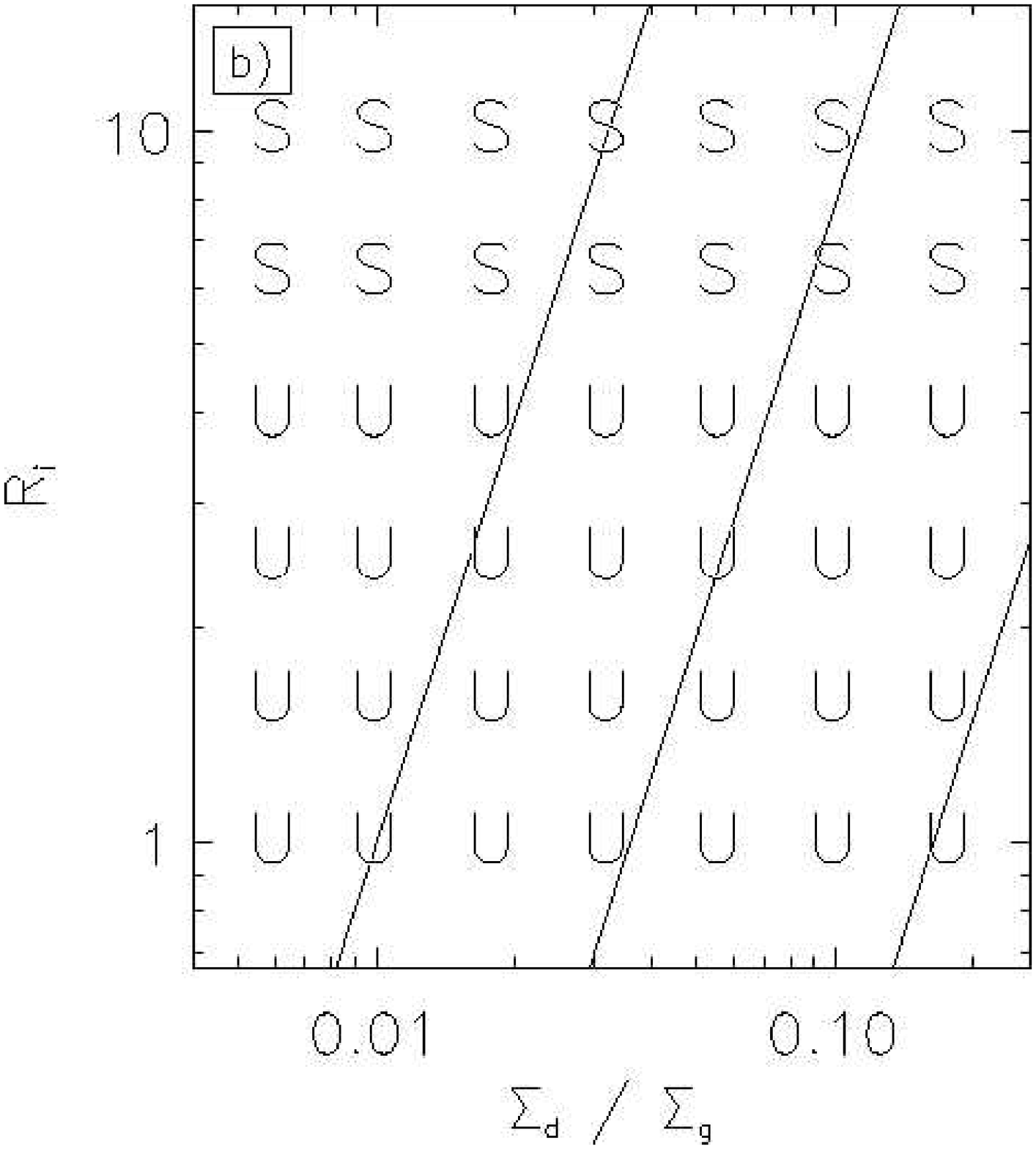}
\parbox{5.5in}{
\caption{
\small
Stability in the dust layer for a matrix of values of the
dust surface density
$\Sigma_d / \Sigma_g$ and Richardson number values $\Ri$,
({\it a}) with Coriolis terms turned off ($\Omega_F=0$) and
({\it b}) with Coriolis terms turned on  ($\Omega_F \neq 0$).
Stable cases, based on simulation growth rates,
are marked with an ``S'' and unstable cases with a ``U.''
Models with a measured growth rate $\omega_I< 0.05 \Omega_K$
are considered stable.
The lines across these plots trace the value of the midplane dust
fraction, (from the left) $\mu=0.3,1.0$ and 10.
Note the different ranges of $\Ri$ in ({\it a}) and ({\it b}).
\bigskip
\hrule
\label{us_fig}
}}
\end{figure}

In Figure \ref{us_fig} we present results for an array of values in
the $\Sigma_d / \Sigma_g - \Ri$ parameter space, noting whether
each model is stable (S) or unstable (U) according to the above
criteria.
Generally speaking, the $\Omega_F=0$ case shows stability for
$\Ri > 0.3$
(Figure \ref{us_fig}a).
Similarly to \S \ref{mu_sec}, we also forced the development
of the odd mode of the instability
by placing a reflecting boundary at $z=0$.
For the odd modes,
the edge of stability also appears consistent with the
criterion $\Ri=1/4$, although the measured growth rates are
much lower than for the even mode.

\begin{figure}[t]
\epsscale{0.5}
\plotone{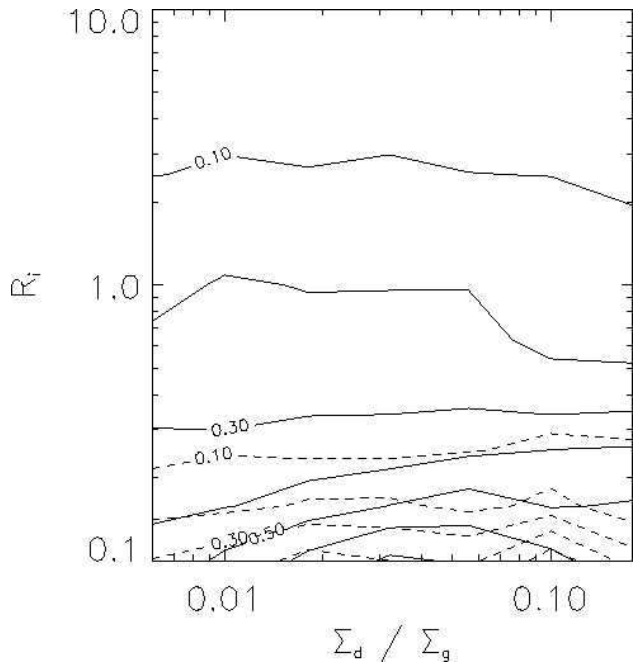}
\parbox{5.5in}{
\caption{
\small
Comparison of growth rates for the $\Omega_F=0$ (dashed lines)
and $\Omega_F \neq 0$ (solid lines) cases, in units of $\Omega_K$.
The growth rates are generally higher when Coriolis terms are
turned on.
\bigskip
\hrule
\label{growth_fig}
}}
\end{figure}

The main difference between the $\Omega_F=0$ and $\Omega_F \neq 0$
cases (Figure \ref{us_fig}b)
is that stability in the $\Omega_F \neq 0$ case requires a much
larger value of the Richardson number, near $\Ri \sim 5$.
It is important to stress that this threshold value for $\Ri$
depends on the value chosen for the fiducial minimum ``unstable''
growth rate, since there is some subjectivity in the
procedure to measure growth rates and
the slope of $\omega_I$ as function of $\Ri$ is quite shallow when
$\Omega_F \neq 0$ (see Figure \ref{growth_fig}).
Nevertheless, the layer is quite clearly unstable for values
of $\Ri$ significatively larger that $1/4$.
Since $\Ri$ can be thought as a proxy for the layer thickness
(eq. [\ref{zmax_eq}]), the condition that
a larger $\Ri$ is required for stability when $\Omega_F \neq 0$
is consistent with the results of \S \ref{mu_sec}, in which we found
that the final dust distribution is thicker when the Coriolis terms
are turned on.

As a coda for this section, we comment on the constant $\Ri$
distribution adopted here.
\citet{sek98} studied the profile determined by a constant-$\Ri$ dust
distribution and noticed that the space density develops an infinite
cusp if the $\Sigma_d / \Sigma_g$ value is large enough.
\citet{you02} interpreted this result as a hint that the gas
distribution can only sustain the weight of a mass of dust similar
to its own mass, and if surpassed, the dust would precipitate.
By numerically integrating our equation (\ref{muapprox_eq}) for a
given $\Ri=const.$, together with the
hydrostatic equilibrium condition in equation (\ref{rhog_eq}), we
find that

\begin{equation}
\mu_0 = \frac{\me}{2}\;
  \exp \left[ \sqrt{\frac{\pi}{2}}\;\frac{\Sigma_d}{\Sigma_g}
  \;\frac{1}{\Ri^{1/2} v_{0,max}/c_s} \right]
\end{equation}

\noindent
for $\mu_0 \gg 1$ (see also eq. [22] in Sekiya 1998).
This means that, for $\Ri_{crit}$ of order unity, $\Sigma_d /
\Sigma_g > v_{0,max}/c_s$ implies an exponential growth of the
midplane dust abundance.
The onset of a cusp in density would therefore occur (under the
condition $\Ri = const.$), for a range of radii $\sim 1-10 \au$ (see
eq. [\ref{v0maxR_eq}]), when $\Sigma_d / \Sigma_g$ exceeds $\sim
0.1$, i.~e., for roughly an order of magnitude dust enhancement.
However, this conclusion appears to be an artifact of the
$\Ri=const.$ condition.
If $\Ri$ is permitted to vary with height, with $\Ri$ becoming large
near the midplane, then $\mu_0$ varies only linearly rather than
exponentially with $\Sigma_d / \Sigma_g$, and a cusp need not form.


\section{SUMMARY AND CONCLUSIONS}
\label{conclusions_sec}

In this work, we have studied the stability of gaseous disks with
vertically varying dust abundances and accompanying vertical velocity
gradients.
We present both linear analysis of a simple discrete three-layer
distribution, and full numerical simulations of a pair of
continuous distributions.
We focus on identifying and intercomparing criteria for instability
when the Coriolis force terms are considered
($\Omega_F \neq 0$) or disregarded ($\Omega_F = 0$).
Our results indicate that the Richardson number $\Ri$
is still a good general discriminant for the stability/instability boundary
when $\Omega_F \neq 0$,
but the critical value is larger than the classical result $\Ri=1/4$
for $\Omega_F=0$.

As with the well known two-layer problem, the critical $\Ri$
for the discrete three-layer distribution
depends on the wavelength of interest.
We argue that wavelengths of order of a few times the layer thickness
($k_y H_d \approx 1$) are the most relevant, since smaller
wavelengths will yield less dust mixing in the non-linear regime,
and larger ones have smaller growth rates.
In addition, for realistic vertical distributions with continuous
variations of $v_y(z)$ and $\mu(z)$, waves with $k_y H_d \gg 1$ would
not be unstable.
For $k_y H_d = \pi/3$, $\Ri_{\eff} < 1.1$ is a necessary (but not
sufficient) condition for instability for the $\Omega_F \neq 0$
case.
In general, when $\Omega_F \neq 0$, marginally-unstable modes first
appear when $\Ri_{\eff}$ drops below $(k_y H_d)^2$.
Since $\Ri_{\eff}$ is a proxy for the layer thickness,
this means that, when Coriolis forces are considered,
the dust layer is thicker than in the $\Omega_F=0$ case
at the onset of instability.
At $\Sigma_d / \Sigma_g = 0.01$, the increase in thickness is
$\sim 50\%$, whereas for $\Sigma_d / \Sigma_g = 0.1$, the increase
is nearly a factor of three.
For the $v_{0,max}/c_s=0.1$ value adopted throughout this work,
$R = 3.5 \au$ (for MSN parameters), and GI will set in only when
$\mu > 187$ \citep{sek98}.
The thicker layer at the ``edge'' of KHI implies that the
dust-to-gas surface density ratio necessary for GI (neglecting
vertical self-gravity) would increase from the already high value of
$\Sigma_d / \Sigma_g = 1.1$ (eq. [\ref{instb_eq}]) to
$\Sigma_d / \Sigma_g = 15$  (eq. [\ref{rotinsb_eq}]), when Coriolis
effects are included.%
\footnote{
  For the continuous dust distribution studied by \citet{gar04}, the
  surface density needed for GI, with these parameters and neglecting
  Coriolis forces, would be $\Sigma_d / \Sigma_g \approx 0.4$, only
  slightly lower than the value 1.1 for our 3 layer model with
  $\Omega_F=0$.
}
While we certainly do not believe that our results (assuming full
coupling) can be accurately extrapolated into the dust-dominated
regime, this comparison gives a sense of the strong change that can
be expected.
The astronomical implication is, of course, that GI is even more
difficult to achieve than previously thought.

We note that consideration of even-symmetry (in $v_z$, i.~e.
midplane-crossing) modes is important for this problem, although
some previous analytic studies \citep[for example]{ish02, ish03}
have focused only on odd-symmetry modes.
The even-symmetry case becomes unstable at a larger value of
$\Ri_{\eff}$ than the odd-symmetry case [which is similar to the
two-layer system analyzed by \citet{cha61}].

The results from our simulations are consistent with the main
conclusions drawn from the simple three-layer stability analysis.
Based on an array of simulations performed that were initialized
with $\Ri=const.$ vertically, we found that
$\Ri \sim 5$ is the critical value
found for the $\Omega_F \neq 0$ case.
This critical $\Ri$ value should be considered an approximate
result, since the procedure used to evaluate stability from the
simulations involves a certain level of subjectivity.
Nevertheless, the dust layer is clearly unstable for $\Ri$ values
significantly larger than $1/4$.
As with the discrete case, the larger $\Ri$ value also implies a
thicker dust layer;
for $\Sigma_d / \Sigma_g =0.01$
at the threshold of instability for $\Omega_F \neq 0$, the height
containing half the dust surface density increases by a factor
$\sim 3$ compared to the corresponding non-rotating model.
Of course, both this and other results based on the strong-coupling
assumption should be considered carefully when $\mu \gg 1$, i.~e.
for large $\Sigma_d / \Sigma_g$.

\citet{you02} point out that precipitation of solids is likely to
occur in layers for which $\mu \gg 1$.
For vertical profiles with $\Ri=const.$, this situation arises first
at the midplane, where a density cusp forms for $\Sigma_d /
\Sigma_g$ exceeding some critical value (of order $v_{0,max}/c_s$).
They therefore conclude that, for GI to develop,
$\Sigma_d / \Sigma_g$ needs to be increased from the MSN value
by a smaller amount than estimates without a density cusp.
Since our results imply that stability fails at larger $\Ri$ (and
hence a thicker dust layer) than the value assumed by \citet{you02},
the total dust enhacement $\Sigma_d / \Sigma_g$ required before a
cusp is predicted (and precipitation develops) would nominally be
larger than they concluded.
However, it is not clear that the condition for cusp formation can
in fact be obtained and written in a simple way in terms of
$\Sigma_d / \Sigma_g$, for the reason we now discuss.

States which have $\Ri=const.$ have been considered of particular
interest because they are everywhere at a nominal threshold for
instability.
However, the proposal that profiles should evolve to such states
implicitly depends on an assumption that
the localized failure of the KH stability criterion would lead
to localized turbulence.
While we chose a constant-$\Ri$ setup in the models of \S
\ref{rich_sec} to facilitate comparison with previous work, the
results of the simulations in fact call this implicit assumption
into question.
When exploring $\Ri(z)$ for the simulations in \S \ref{mu_sec},
in some of our models the $\Ri > 1/4$ stability condition failed
only for $|z| < 2 H_d$.
Nevertheless, the fastest growing mode in these models
had $\lambda \approx 7 H_d$,
and the non-linear evolution involved the whole layer,
yielding dust mixing at heights over 10 times the
initial $H_d$ value.
Furthermore, when the simulations were followed until the dust
distribution was somewhat homogeneous in the azimuthal direction,
the vertical profile did not appear to have a constant $\Ri$ value.

It is our view that, as exemplified by these models,
a local failure to satisfy a stability criterion which is stated in
terms of local quantities need not
trigger only local instabilities, i.~e. of wavelengths similar to
the thickness of the ``failing'' region.
Instead, the wavelength will be similar to
the overall velocity gradient, and the height of the waves when they
break will be similar to the wavelength.
The re-adjustment of the dust distribution that is generated will
thus be {\it global}, including regions that initially have
$\Ri > \Ri_{crit}$.
It is, therefore, not obvious
that the ``final,'' or quasi-equilibrium, distribution should be
accurately described by a local quantity.
It may be that the marginal state that develops as a result of the
competition between dust settling and shear-driven turbulence in
fact has a relatively uniform distribution of dust in a central
layer, rather than a cusp out of which solids can precipitate.
Of course, if particles grow large enough prior to the onset of
global dust-layer turbulence, then they may continue to settle
faster than the growth rate of the instabilities.
Since particle growth depends on uncertain sticking efficiency,
however, this question remains open.
Observations with newly deployed infrared telescopes should shed
some light on the true vertical distribution of solids in
proto-planetary disks.

For the technical modeling in this work, we have adopted a number of
simplifications and idealizations.
While we believe that our main conclusions are not sensitive to
these assumptions, it is appropiate to review what they are.
One such simplification is that we consider only models with
$k_x=0$, and neglect radial shear and tidal forces.
We focused on $k_x=0$ because these modes typically have the fastest
growth rates and largest instability thresholds in terms of $\Ri$.
Had we included shear, then $k_x$ would have grown in time as
$\dif k_x/\dif t = (3/2) \Omega_K k_y$, which would ultimately limit
growth of perturbations of a given $k_y$.
In future work, it will be important to examine how, for example,
the total amplification of shearing KH wavelets depends on $\Ri$.

Other idealizations include neglecting the slip of gas relative to
dust, treating the large-scale radial gradient of pressure as fixed
in time, and neglecting other sources of turbulence.
Gas-dust slip can itself lead to streaming instabilities
\citep{you05}.
However, either with or without full coupling of the gas and dust,
the instabilities that develop depend on the gas having consistent
sub-keplerian azimuthal velocities.
This relies on having a consistent radial pressure gradient;
in principle, this condition could be invalidated if there are
sufficient pressure perturbations induced by turbulence or
large-scale waves in the disk.
The ratio of such terms, $\sim (\delta P \, R) / (P_0 \, \lambda)$,
could be significant even for moderate amplitude perturbations with
$\lambda \sim H_g$.
In future work, we intend to explore these and other dynamical
processes that may affect the settling and mixing of solids in dusty
disks.


\acknowledgements

We gratefully ackowledge stimulating discussions with Jim Stone,
Jeremy Goodman, and Marco Martos,
and helpful comments on the manuscript by Andrew Youdin, Jeremy
Goodman, and an anonymous referee.
The simulations presented in this paper were carried out
using the Beowulf cluster administered by the Center for
Theory and Computation of the Department of Astronomy at the
University of Maryland.
Financial support was provided by NASA grant NAG511767.


\appendix

\section{SOLUTION TO THE DISPERSION RELATION, FOR $\Omega_F \neq 0$}
\label{app_sec}

The dispersion relation for the even-symmetry mode, 
equation (\ref{disprel_eq}), in the long-wave\-length regime, is
given in equation (\ref{disprel_long_eq}).
To investigate the nature of the solutions of this non-algebraic
equation, we follow a graphical procedure similar to that in
\citet[p. 498]{cha61}.
Define

\begin{eqnarray}
\xi  &\equiv& \frac{\omega}{\Omega_K} \\
\eta &\equiv& \frac{\omega - k_y V}{\Omega_K}
      =  \frac{\omega}{\Omega_K} -
  \frac{\mu}{1+\mu} \, \frac{H_g}{H_d} \,
  \frac{v_{0,max}}{c_s} \, k_y H_d \\
\beta &\equiv& \frac{2 \Omega_F}{\Omega_K}.
\end{eqnarray}

\noindent
In these variables, the dispersion relation reads

\begin{equation}
\eta^2 = \frac{\mu}{1+\mu} \left[ 1 - \frac{\xi^2}{\mu k_y H_d}
  \left(1 - \frac{\beta^2}{\xi^2} \right)^{1/2} \right].
\label{etam_eq}
\end{equation}

\noindent
By simultaneously considering the equation

\begin{equation}
\eta^2 = \frac{\mu}{1+\mu} \left[ 1 + \frac{\xi^2}{\mu k_y H_d}
  \left(1 - \frac{\beta^2}{\xi^2} \right)^{1/2} \right],
\label{etap_eq}
\end{equation}

\noindent
together with the relation $|\xi-\eta| = k_y V / \Omega_K = const.$,
we can obtain a fourth order polynomial instead of a non-algebraic
equation.
All the solutions of the polynomial must be solutions either of
equation (\ref{etam_eq}) or (\ref{etap_eq}).

\begin{figure}[t]
\epsscale{0.5}
\plotone{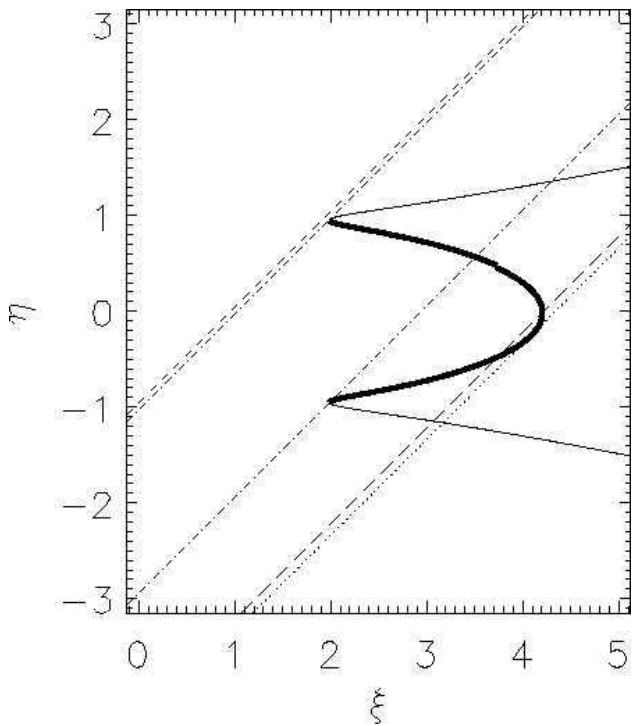}
\parbox{5.5in}{
\caption{
\small
Locus of the real solutions to equations (\ref{etam_eq})
(thick solid line) and (\ref{etap_eq}) (thin solid line),
for $k_y H_d=\pi/3$ and $\mu=10$.
Straight lines $\xi - \eta=const.$ are related to
critical points of the problem (see text).
For given values of $k_y H_d$, $\mu$ and $v_{0,max}/c_s$,
varying
values of $\xi - \eta$ correspond to different values for
$H_d / H_g$.
\bigskip
\hrule
\label{chandra_plot_fig}
}}
\end{figure}

The thick solid line in
Figure \ref{chandra_plot_fig} shows the locus of the real solutions of
equation (\ref{etam_eq}), which we shall call the $P$-branch,
while the thin solid line shows the real solutions of equation
(\ref{etap_eq}), which we shall call the $S$-branch.
For clarity, we show only the $\xi>0$ side of the diagram.
In addition, Figure \ref{chandra_plot_fig} shows several lines
representing $|\xi - \eta| = const$, which relate to
a series of critical points $(\xi,\eta)$ for the problem as follows:

\begin{enumerate}

\item
  The ends of the $P$-branch lie at
  $(\pm \beta, \pm \eta_0)$; these are the intersections with the
  dash-dotted lines $\xi - \eta = \beta \pm \eta_0$ in Figure
  \ref{chandra_plot_fig}.
  Here, $\eta_0^2 = \mu / (1+\mu)$.

\item
  The maximum extention of $|\xi|$ for the $P$-branch lies at
  $(\pm \xi_0,0)$; this is the intersection with the long-dashed
  line $\xi - \eta = \xi_0$ in Figure \ref{chandra_plot_fig}.
  Here, $\xi_0^2=\sqrt{(\mu k_y H_d)^2 + \beta^4/4}+\beta^2/2$.

\item
  The point of the $P$-branch with the maximum value of
  $|\xi - \eta|$, $(\xi_t, \eta_t)$;
  this is at the intersection with the dotted line in Figure
  \ref{chandra_plot_fig}.

\item
  The $(\xi, \eta) = (0,-\eta_0)$ point, where $\eta_0^2= \mu/(1+\mu)$;
  the short-dashed line $\xi - \eta = -\eta_0$
  in Figure \ref{chandra_plot_fig} runs through this point.

\end{enumerate}

Any of the four solutions to the equivalent fourth-order polynomial
must be a solution of either
equation (\ref{etam_eq}) or (\ref{etap_eq}),
and the number or nature of the roots of each equation can change only
at the critical points outlined above.
The roots corresponding to equation (\ref{etam_eq}) are shown in
Figure \ref{chandra_fig}, for an array of $H_d / H_g$
(or equivalently $|\xi-\eta|$) values, given $\mu=1$ (a), and
$\mu=10$ (b), and $k_y H_d=\pi/3$.

On the left the diagrams in Figure \ref{chandra_fig}, at
small $H_d / H_g$ (or on the right of Figure \ref{chandra_plot_fig},
at large $|\xi - \eta|$), there is a pair of complex roots.
In Figure \ref{chandra_fig}b, these two complex roots become real as
$|\xi-\eta|$ becomes less than $\xi_t - \eta_t$ since then
a $|\xi-\eta|=const.$ line crosses the $P$- and $S$-branches four
times, implying all four roots are real.
Moving further right in Figure \ref{chandra_fig}b (or left in Figure
\ref{chandra_plot_fig}),
one of those roots becomes complex when $|\xi-\eta| < \beta+\eta_0$,
and the other real root disappears when $|\xi-\eta| < \beta-\eta_0$.
In Figure \ref{chandra_fig}a, at lower $\mu$, a complex root is
present whenever a real root exists.
At the far right in Figures \ref{chandra_fig}a and
\ref{chandra_fig}b (corresponding to the region of Figure
\ref{chandra_plot_fig} with $|\xi-\eta| < \eta_0$), there are no
solutions.

There is a range of values of $|\xi-\eta|$ with all real
solutions (implying stability for that set of parameters)
if the $P$-branch has a section with $|\dif\eta/\dif\xi|<1$.
When that happens (as for the $\mu=10$ case of Figure
\ref{chandra_fig}b), the points of the $P$-branch with
$|\dif\eta/\dif\xi|=1$ lie very close to the $(\beta,-\eta_0)$ and
$(\xi_0,0)$ points.
So, a good approximation for the opening of such stability gap is
$\xi_0 > \beta+\eta_0$.
While it is straightforward to find a precise criterion for the
existence of this gap,
it is of limited practical significance since the gap closes for
longer wavelengths (which yield more efficient vertical mixing).
We consider more significant the fact that
there are no solutions, real or complex, for $|\xi-\eta|<\eta_0$,
since this yields a necessary (but not sufficient) condition for
instability.



\begin{thebibliography}


\bibitem[Champney, Dobrovolskis \& Cuzzi(1995)]{cha95}
  Champney, J. M., Dobrovolskis, A. R. \& Cuzzi, J. N.
  1995, Phys. Fluids, 7, 1703

\bibitem[Chandrasekhar(1981)]{cha61}
  Chandrasekhar, S.
  1981, Hydrodynamic and Hydromagnetic Stability
  (New York: Dover)

\bibitem[Coradini, Federico \& Magni(1981)]{cor81}
  Coradini, A., Federico, C. \& Magni, G.
  1981, \aap, 98, 173

\bibitem[Cuzzi, Dobrovolskis \& Champney(1993)]{cuz93}
  Cuzzi, J. N., Dobrovolskis, A. R. \& Champney, J. M.
  1993, Icarus, 106, 102

\bibitem[Cuzzi, Dobrovolskis \& Hogan(1994)]{cuz94}
  Cuzzi, J. N., Dobrovolskis, A. R. \& Hogan, R. C.
  1994, Lunar Planet. Sci. XXV, 307

\bibitem[Drazin \& Reid(1981)]{dra81}
  Drazin, P. G. \& Reid, W. H.
  1981, Hydrodynamic Stability
  (Cambridge: Cambridge University Press)

\bibitem[Garaud \& Lin(2004)]{gar04}
  Garaud, P. \& Lin, D. N. C.
  2004, \apj, 608, 1050

\bibitem[Goldreich \& Ward(1973)]{gol73}
  Goldreich, P. \& Ward, W. R.
  1973, \apj, 183, 1051

\bibitem[Haghighipour \& Boss(2003)]{hag03}
  Haghighipour, N. \& Boss, A. P.
  2003, \apj, 583, 996

\bibitem[Hayashi(1981)]{hay81}
  Hayashi, C.
  1981, Prog. Theor. Phys. Suppl., 70, 35

\bibitem[Howard(1961)]{how61}
  Howard, L. N.
  1961, J. Fluid Mech., 10, 509

\bibitem[Huppert(1968)]{hup68}
  Huppert, H. E.
  1968, J. Fluid Mech., 33, 353

\bibitem[Ishitsu \& Sekiya(2002)]{ish02}
  Ishitsu, N. \& Sekiya, M.
  2002, Earth Planets Space, 54, 917

\bibitem[Ishitsu \& Sekiya(2003)]{ish03}
  ---
  2003, Icarus, 165, 181

\bibitem[Li, Goodman \& Narayan(2003)]{li03}
  Li, L.-X., Goodman, J., \& Narayan, R.
  2003, \apj, 593, 980

\bibitem[Miles(1961)]{mil61}
  Miles, J. W.
  1961, J. Fluid Mech., 10, 496

\bibitem[Nakagawa, Nakazawa \& Hayashi(1981)]{nak81}
  Nakagawa, Y., Nakasawa, K. \& Hayashi, C.
  1981, Icarus, 45, 517

\bibitem[Nakagawa, Sekiya \& Hayashi(1986)]{nak86}
  Nakagawa, Y., Sekiya, M. \& Hayashi, C.
  1986, Icarus, 67, 355

\bibitem[Rice et al.(2004)]{ric04}
  Rice, W. K. M., Lodato, G., Pringle, J. E., Armitage, P. J.
  \& Bonnell, I. A.
  2004, \mnras, 355, 543

\bibitem[Safranov(1969)]{saf69}
  Safranov, V. S.
  1969, Evolution of the Protoplanetary Cloud and the Formation of
  the Earth and Planets
  (Moscow: Nauka Press)

\bibitem[Sekiya(1983)]{sek83}
  Sekiya, M.
  1983, Progress of Theoretical Physics, 69, 1116

\bibitem[Sekiya(1998)]{sek98}
  ---
  1998, Icarus, 133, 298

\bibitem[Sekiya \& Ishitsu(2000)]{sek00}
  Sekiya, M. \& Ishitsu, N.
  2000, Earth Planets Space, 52, 517

\bibitem[Sekiya \& Ishitsu(2001)]{sek01}
  ---
  2001, Earth Planets Space, 53, 761

\bibitem[Stone \& Norman(1992)]{sto92}
  Stone, J. M. \& Norman, M. L.
  1992, \apjs, 80, 753

\bibitem[Tanga et al.(2004)]{tan04}
  Tanga, P., Weidenschilling, S. J., Michel, P. \&
  Richardson, D. C.
  2004, \aap, 427, 1105

\bibitem[Weidenschilling(1977)]{wei77}
  Weidenschilling, S. J.
  1977, \mnras, 180, 57

\bibitem[Weidenschilling(1980)]{wei80}
  ---
  1980, Icarus, 44, 172

\bibitem[Weidenschilling(1995)]{wei95}
  ---
  1995, Icarus, 116, 433

\bibitem[Weidenschilling(2003)]{wei03}
  ---
  2003, Icarus, 165, 438

\bibitem[Weidenschilling \& Cuzzi(1993)]{wei93}
  Weidenschilling, S. J. \& Cuzzi, J. N.
  1993, in Protostars and Planets III,
  ed. E. H. Levy \& J. I. Lunine
  (Tucson: University of Arizona Press), 1031

\bibitem[Youdin \& Shu(2002)]{you02}
  Youdin, A. N. \& Shu, F. H.
  2002, \apj, 580, 494

\bibitem[Youdin \& Chiang(2004)]{you04}
  Youdin, A. N. \& Chiang, E. I.
  2004, \apj, 601, 1109

\bibitem[Youdin \& Goodman(2005)]{you05}
  Youdin, A. N. \& Goodman, J.
  2005, \apj, 620, 459



\end{thebibliography}
\end{document}